\documentclass[journal, onecolumn,12pt, draftclsnofoot]{IEEEtran}

\usepackage{mathrsfs}
\usepackage[noadjust]{cite}
\usepackage{graphicx,color,overpic,psfrag}
\usepackage{float}
\usepackage{stfloats}
\usepackage{amsmath, amssymb}
\usepackage{cleveref}
\usepackage{latexsym}
\usepackage{bm}
\usepackage{amssymb}
\usepackage{cases}
\usepackage{array}
\usepackage{fancyhdr}
\usepackage{setspace}

\ifCLASSOPTIONcompsoc
\usepackage[caption=false,font=normalsize,labelfont=sf,textfont=sf]{subfig}
\else
\usepackage[caption=false,font=footnotesize]{subfig}
\fi

\definecolor{myblue}{rgb}{.91,.95,.99}
\usepackage{multirow}
\usepackage{url}
\usepackage{algorithm,algpseudocode,float}

\usepackage{blkarray}
\usepackage{booktabs}

\usepackage{multirow}
\usepackage{dsfont}
\usepackage{tabularx}
\usepackage[table]{xcolor}
\usepackage{amsfonts}

\usepackage{letltxmacro}

\graphicspath{{figure/}}

\newcolumntype{L}{>{\hspace*{-\tabcolsep}}l}
\newcolumntype{R}{c<{\hspace*{-\tabcolsep}}}
\definecolor{lightblue}{rgb}{0.93,0.95,1.0}

\newtheorem{lemma}{Lemma}

\newcommand{\lmref}[1]{Lemma \ref{#1}}

\newcommand{\figref}[1]{Fig. \ref{#1}}

\newcommand{\alref}[1]{Algorithm \ref{#1}}

\newcommand{\secref}[1]{Section \ref{#1}}

\newcommand{\Exp}{{\mathsf{E}}}
\newcommand{\expect}[1]{\Exp\left\{#1\right\}}

\newcommand{\tr}[1]{\mathsf{tr}\left\{#1\right\}}

\newcommand{\thetabs}[2]{{\dnnot{\theta}{bs}}}

\newcommand{\cC}{\mathcal{C}}

\newcommand{\cO}{\mathcal{O}}
\newcommand{\cP}{\mathcal{P}}
\newcommand{\cQ}{\mathcal{Q}}

\newcommand{\cS}{\mathcal{S}}

\newcommand{\ba}{\mathbf{a}}
\newcommand{\bb}{\mathbf{b}}
\newcommand{\bc}{\mathbf{c}}

\newcommand{\bg}{\mathbf{g}}

\newcommand{\bn}{\mathbf{n}}

\newcommand{\bp}{\mathbf{p}}
\newcommand{\bq}{\mathbf{q}}

\newcommand{\bx}{\mathbf{x}}
\newcommand{\by}{\mathbf{y}}
\newcommand{\bz}{\mathbf{z}}

\newcommand{\bA}{\mathbf{A}}
\newcommand{\bB}{\mathbf{B}}
\newcommand{\bC}{\mathbf{C}}
\newcommand{\bD}{\mathbf{D}}
\newcommand{\bE}{\mathbf{E}}

\newcommand{\bG}{\mathbf{G}}
\newcommand{\bH}{\mathbf{H}}
\newcommand{\bI}{\mathbf{I}}

\newcommand{\bM}{\mathbf{M}}
\newcommand{\bN}{\mathbf{N}}

\newcommand{\bQ}{\mathbf{Q}}
\newcommand{\bR}{\mathbf{R}}
\newcommand{\bS}{\mathbf{S}}
\newcommand{\bT}{\mathbf{T}}

\newcommand{\bW}{\mathbf{W}}
\newcommand{\bX}{\mathbf{X}}

\newcommand{\bZ}{\mathbf{Z}}

\newcommand{\C}{\mathbb{C}}
\newcommand{\R}{\mathbb{R}}

\newcommand{\bzero}{\mathbf{0}}

\newcommand{\dnnot}[2]{#1_{\mathrm{#2}}}

\newcommand{\ntb}{\notag\\}

\newcommand{\figsincolwid}{12cm}

\allowdisplaybreaks

\begin{document}

\title{\huge Near-Field Wideband Extremely Large-scale MIMO Transmission with Holographic Metasurface Antennas}

\author{
Jie~Xu,
Li~You,
George~C.~Alexandropoulos,
Xinping~Yi,
Wenjin~Wang,
and~Xiqi~Gao

\thanks{
Jie~Xu, Li~You, Wenjin~Wang, and Xiqi~Gao are with the National Mobile Communications Research Laboratory, Southeast University, Nanjing 210096, China, and also with the Purple
Mountain Laboratories, Nanjing 211100, China (e-mail: xujie@seu.edu.cn; lyou@seu.edu.cn; wangwj@seu.edu.cn; xqgao@seu.edu.cn).
}
\thanks{
George~C.~Alexandropoulos is with the Department of Informatics and Telecommunications, National and Kapodistrian University of Athens, Panepistimiopolis Ilissia, 15784 Athens, Greece and also with the Technology Innovation Institute, 9639 Masdar City, Abu Dhabi, United Arab Emirates (e-mail: alexandg@di.uoa.gr).
}
\thanks{
Xinping~Yi is with the Department of Electrical Engineering and Electronics, University of Liverpool, Liverpool L69 3BX, U.K. (e-mail: xinping.yi@liverpool.ac.uk).
}
}
\maketitle

\begin{abstract}
Extremely large-scale multiple-input multiple-output (XL-MIMO) is the development trend of future wireless communications. However, the extremely large-scale antenna array could bring inevitable near-field and dual-wideband effects that seriously reduce the transmission performance.
This paper proposes an algorithmic framework to design the beam combining for the near-field wideband XL-MIMO uplink transmissions assisted by holographic metasurface antennas (HMAs). Firstly, we introduce a spherical-wave-based channel model that simultaneously takes into account both the near-field and dual-wideband effects. Based on such a model, we then formulate the HMA-based beam combining problem for the proposed XL-MIMO communications, which is challenging due to the nonlinear coupling of high dimensional HMA weights and baseband combiners. We further present a sum-mean-square-error-minimization-based algorithmic framework. Numerical results showcase that the proposed scheme can effectively alleviate the sum-rate loss caused by the near-field and dual-wideband effects in HMA-assisted XL-MIMO systems. Meanwhile, the proposed HMA-based scheme can achieve a higher sum rate than the conventional phase-shifter-based hybrid analog/digital one with the same array aperture.
\end{abstract}


\section{Introduction}\label{sec:Introduction}
Extremely large-scale multiple-input multiple-output (XL-MIMO), also referred to as ultra-massive MIMO, is a promising technology to support future wireless communications \cite{2015Massive, 2022CuiNear, 2022Zhang6G, 2020WirelessTang}. However, there remain some inevitable problems when conventional beamforming techniques are adopted for realization. In conventional fully digital transceivers, each antenna element is connected to an individual radio frequency (RF) chain, which is composed of a power amplifier, a digital filter, an analog-to-digital converter (ADC), and a mixer. Since the RF chain is expensive and ADCs dominate the total power consumption, simply increasing the conventional antenna elements will result in high power consumption and hardware cost, as well as the huge physical size.

Holographic metasurface antenna (HMA), also named as dynamic metasurface antenna, has been proposed as a brand-new antenna paradigm in recent years. The metamaterial elements, the core technology behind HMA, can be distributed with sub-wavelength intervals, so a massive number of metasurface antennas can be integrated into a compact space to generate a holographic MIMO array \cite{2020Holographic,2020HolographicPizzo,2021HMA}.
Since each element of HMAs contains its own switchable components, such as positive intrinsic-negative diodes, the effective parameters of each element, especially their permittivity and permeability \cite{2004Metamaterials}, 
can be independently tailored to achieve the desired signal response. An extremely large-scale HMA-based array can be obtained by tiling together multiple waveguides, whose top layers are embedded with several metamaterial elements. Such an array is connected to a digital processor through separate input/output ports, providing the ability of controllable and power-efficient beam forming/combining in real-time \cite{2021Dynamic6G}. When working on the receiving side, HMAs capture signals from channels, process them in the analog domain, and transmit them along the waveguides to the digital processor. Correspondingly, on the transmitting side, signals generated by digital processors pass through input ports to HMAs and then are radiated to channels by metamaterial elements.

As a brand-new antenna paradigm, HMAs have some unique advantages over traditional antennnas. Firstly, the needed RF chains in an HMA-based transceiver can be much fewer than those in a conventional fully digital transceiver. Specifically, the number of RF chains is generally equal to that of waveguides in an HMA-based transceiver, which is much smaller than that of metamaterial elements. As a result, the power consumption can be significantly reduced and the energy efficiency (EE) is obviously improved \cite{2021DMAsEE}.
Secondly, metamaterial elements are regularly arranged at sub-wavelength intervals on the waveguides, which indicates that an HMA-based array can accommodate more antenna elements than a conventional array, e.g., a patch antenna array, with the same aperture. 
By contrast, for the conventional array, antenna elements are usually distributed with half wavelength to reduce the mutual coupling, leading to huge sizes of antenna arrays in XL-MIMO systems \cite{2022ZhangHISAC}, \cite{2021GradoniEnd}. 
Thus, HMAs are more conducive to the layout of large-scale antenna arrays compared to conventional antennas, especially in the circumstance of a limited area. Finally, HMA-based transceivers can implement a form of hybrid analog/digital (A/D) beamforming with hardware and power savings. In particular, the conventional hybrid A/D beamforming is usually supported by a large number of active phase shifters \cite{2016PhaseorSwitches}, which brings non-negligible power consumption. However, HMAs utilize power-saving holographic techniques to inherently accomplish signal processing in the analog domain without any additional circuits \cite{2021Dynamic6G}. Therefore, HMA-based beamforming is more power-efficient than the hybrid A/D one. 

As mentioned above, future wireless communications tend to adopt the extremely large-scale antenna array at the base station (BS) for sheer number of user terminals. As a result, the near-field region of the BS is enlarged, which can reach several dozens or even hundreds of meters, making it possible for wireless communications to take place in the near field \cite{2022CuiNear,2021NearWideDai,2022Zhang6G}. Generally, the wavefront of an electromagnetic wave is approximated to a plane for far-field communication, which means the channels' array steering vectors only depend on the angle of arrival/departure. However, when wireless communications take place in the near-field region, the signal angles of arrival/departure cannot be approximately equal for all antenna elements. Instead, the array steering vectors should be characterized by both angles of arrival/departure and distances between scatters and the BS, otherwise it will cause significant loss to the sum-rate performance \cite{2022NearField}. This phenomenon is called the \emph{near-field effect}. Meanwhile, extremely large-scale arrays could give rise to the spatial-wideband and frequency-selective effects \cite{2018Spatialwideband}, \cite{2021ChannelWideband}. In the conventional small-scale array at the BS, the time delay differences among all the receiving antenna elements are generally much smaller than the symbol interval, which can be ignored without loss of  generality. However, for extremely large-scale arrays, the maximum time delay between different antenna elements is very likely to be comparable to or larger than the symbol interval, which results in the unsynchronized reception. In such a case, the antenna elements may receive different symbols at the same time. This phenomenon is called the spatial-wideband effect. Additionally, for a wideband transmission, signals obtain different gains at different frequencies, thus bringing the frequency-selective effect. We use the term \emph{dual-wideband effects} to represent both spatial-wideband and frequency selective effects. With these issues taken into account, a beam forming/combining method is urgently required to alleviate the near-field and dual-wideband effects in HMA-based near-field wideband XL-MIMO communications.

There exist several works on the HMA-based beam forming/combining for wireless communications. The authors in \cite{2019Dynamicuplink} and \cite{2019DynamicDownlink} proposed the transceiver models for HMA-based uplink and downlink systems with multiple single-antenna users, respectively, and preliminarily evaluated the sum-rate performance of the HMA-based systems. The authors in \cite{2021Dynamic6G} proposed the concept of HMA-based precoding and demonstrated its sum-rate performance gain over the fully-connected hybrid A/D precoding. The authors in \cite{2021DMAsEE} studied the EE performance of the HMA-based MIMO uplink system and verified that the proposed EE performance is significantly better than that based on either the conventional fully-digital or the fully-connected hybrid A/D beam combining architectures. The authors in \cite{2022NearField} proposed an HMA-based transceiver model in the near-field region for a narrowband MIMO downlink system and verified HMAs' beam focusing capability to communicate with users in close proximity.
In \cite{2021DynamicOFDM}, an HMA-based quantitative model for orthogonal frequency division modulation (OFDM) receivers was presented and it was numerically shown that HMA-based receivers with limited bits could accurately recover OFDM signals.
Although HMA-based beam forming/combining has been proposed as a promising technique in these scenarios, their applications for XL-MIMO uplink communications in the near-field region have not been investigated yet.

 Motivated by the above concerns, we investigate the HMA-based beam combining of wideband XL-MIMO uplink communications in the near-field region. The contributions of the paper are summarized as follows:
\begin{itemize}
  \item We propose to mitigate both near-field and dual-wideband effects for the HMA-assisted beam combining for the first time. Specifically, we introduce a channel model for single-cell multi-user HMA-based XL-MIMO uplink communications, in which the spherical wave model is used to characterize both the near-field and spatial-wideband effects. Based on this model, we investigate the problem of designing HMA weights and baseband combiners to facilitate HMA-based near-field wideband XL-MIMO communications.
  \item We propose an algorithmic framework to effectively optimize the beam combining in HMA-based XL-MIMO communications, which can alleviate the sum-rate loss caused by near-field and dual-wideband effects. Firstly, we adopt a matrix-weighted sum-mean-square error minimization (MMSE) approach to simplify the complicated beam combining problem with a sum-rate maximization objective function.
      Then, to address the nonlinear coupling between HMA weights and baseband combiners, we adopt an alternating optimization (AO) method to obtain a feasible solution with low complexity. Moreover, we obtain HMA weights by the matrix vectorization and minorization-maximization (MM) approaches to handle the challenges caused by the physical structure and non-convex feasible entries of the HMA weight matrix.
  \item Numerical results showcase that the proposed algorithmic framework can effectively alleviate the near-field and dual-wideband effects in the HMA-based near-field wideband XL-MIMO uplink communications. Moreover, for the same array aperture, the HMA-based beam combining can achieve a better sum-rate performance than the hybrid A/D combining one based on the conventional antennas. It is also demonstrated that the mutual coupling has negligible effect on the sum-rate performance of the HMA-assisted system, even when the antenna interval is sub-wavelength.
\end{itemize}

The rest of this paper is organized as follows. \secref{sec:system model} introduces the HMA model, the channel model containing the near-field and dual-wideband effects for the HMA-based MIMO uplink, and formulates an HMA-based near-field wideband beam combining problem. \secref{sec:mathematical model} proposes an algorithm to design baseband combiners and HMA weights of the proposed system, including the MMSE, AO, matrix vectorization and MM methods. Numerical results are presented in \secref{sec:Simulation}. Finally, we conclude this paper in \secref{sec:Conclusion}.

Some notations are defined as follows. We use boldface upper-case letters to
denote matrices, e.g., $\bM$, and boldface lower-case letters to denote column vectors, e.g., $\bx$; we use $(\bM)_{i,j}$ to denote the $(i,j)$th entry of $\bM$ and $\bx_{i}$ to denote the $i$th entry of $\bx$; the notations $\C$ and $\R$ represent sets of complex numbers and of real numbers, respectively; the superscripts $(\cdot)^H$, $(\cdot)^{-1}$, and $(\cdot)^T$ represent the matrix conjugate-transpose, inverse, and transpose, respectively; the operator $\odot$ denotes Hadamard product; $\tr{\bM}$, $|\bM|$, and $\operatorname{Re}\{ \cdot \}$ denote the matrix trace, matrix determinant, and real part of the input operations, respectively; the operator $|| \cdot ||_{\text{2}}$ means to obtain $\ell_2$-norm of the input; the notation $\lfloor \cdot \rfloor$ is the integer floor function; the notation $\jmath$ denotes the imaginary unit.


%
%
%
\section{System Model}\label{sec:system model}
Our work considers a single-cell multi-user XL-MIMO uplink system. The BS adopts an HMA-based array to simultaneously receiving signals from $U$ single-antenna users distributed in the near-field region. In this section, we introduce the input-output relationship of HMAs, the channel model, and formulate the problem of designing HMA weights and baseband combiners to facilitate XL-MIMO uplink transmission.

\subsection{Holographic Metasurface Antennas}\label{sec:HMA}
\begin{figure}
	\centering
    \includegraphics[width=\figsincolwid]{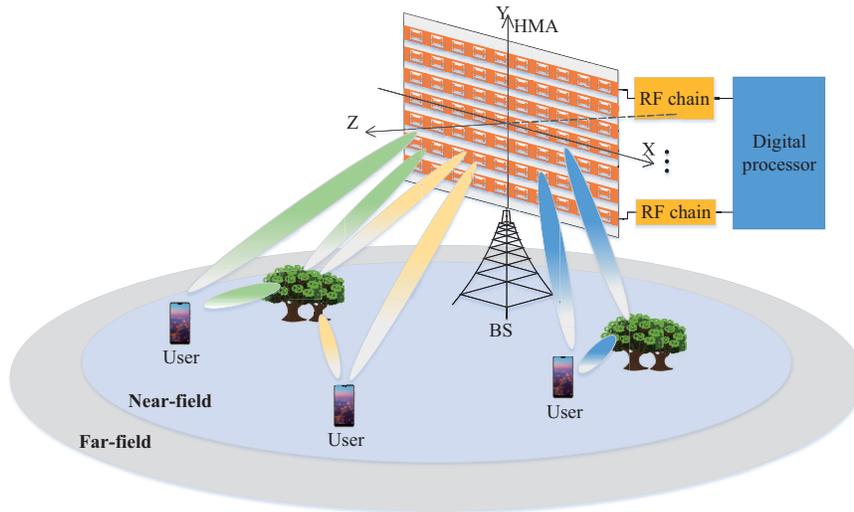}
	\caption{The considered HMAs-assisted XL-MIMO uplink system.}
	\label{fig:cartesian_model}
\end{figure}

The HMA is an emerging concept for the aperture antenna design that utilizes resonant and sub-wavelength metamaterials to implement controllable signal processing in the analog domain \cite{2021Dynamic6G}. In general, metamaterial elements are patterned on the top of a waveguide to form a microstrip transmission line. An extremely large-scale array can thus be generated by increasing the number of microstrips \cite{2021DynamicOFDM}. As shown in \figref{fig:cartesian_model},
we consider an HMA-based array formed by $M$ microstrips, each of which is embedded with $L$ sub-wavelength metamaterial elements. We define the total number of metamaterial elements of the HMA-based array as $N_\text{R} \triangleq M L$.

On the receiving side, metamaterial elements capture signals, adjust amplitudes and phases of signals, and transform them along microstrips to the corresponding RF chains. The output signal of each microstrip is the linear combination of the radiation observed by all metamaterial elements on the microstrip. Such linear combination is related to the following respects. Firstly, metamaterial elements can be treated as resonant electrical circuits. Their frequency responses can be typically modeled as the amplitude-only, binary amplitude, or Lorentzian-constrained phase weights as follows \cite{2017Analysis}, \cite{2016WaveguideFed}:
\begin{itemize}
  \item \emph{Amplitude-only}, i.e., $q \in \cQ= \left[a, b \right]$ where $b \textgreater a \textgreater 0$;
  \item \emph{Binary amplitude}, i.e., $q \in \cQ= c \cdot \{0, 1\}$ where $c \textgreater 0$;
  \item \emph{Lorentzian-constrained phase}, i.e., $q \in \cQ= \{ \frac{\jmath+e^{\jmath \phi}}{2} : \phi \in [0, 2\pi] \}$.
\end{itemize}
Secondly, since each output port is at the edge of the microstrip, signals undergo different time delays propagating from different metamaterial elements to that port. This propagation can be treated as a causal filter with a finite impulse response depending on the wavenumber and element location. We use $\beta$ to denote the wavenumber and $\rho_l$ to denote the factor proportional to the distance between the output port and the $l$th element. Then, the propagation in the frequency domain is proportional to $e^{-\jmath \beta \rho_l}$ \cite{2017Analysis}. 

Typically, an HMA-based array consists of multiple microstrips with one-dimensional planar structures. Let $\mathbf{Q} \in {\mathbb{C}^{M \times N_\text{R}}}$ be a configurable weight matrix that denotes the response of the HMA-based array, which can be formulated as
    \begin{align}\label{eq:Q_constraint}
    \left( \mathbf{Q} \right)_{m_1,(m_2-1)L+l}=
        \left\{
            \begin{aligned}
                & q_{m_1,l},\quad m_1=m_2 \\
                & 0,\quad \ \quad m_1\ne m_2\\
             \end{aligned}
        \right. ,
    \end{align}
where $ m_1 \in {\{1,2,\ldots,M\}}$, $ m_2 \in {\{1,2,\ldots,M\}}$, and $l \in {\{1,2,\ldots,L\}}$. In addition, we let the diagonal matrix $\bH(f) \in \mathbb{C}^{N_\text{R} \times N_\text{R}}$ characterize the frequency-dependent effect of the signal propagation along the microstrips, i.e., $\bH_{l,l}(f) = e^{-\alpha \rho_l - \jmath \beta \rho_l}$, where $\alpha$ is the waveguide attenuation coefficient.

In the HMA-based beam combining receiver, signals are processed by metamaterial elements in the analog domain and by a baseband combiner in the digital domain. Then, the post-process receiving signal can be represented in the frequency domain as
    \begin{align}\label{eq:zf}
    \bz(f) = \bW(f)^H \mathbf{Q} \bH(f) \by(f) \in {\mathbb{C} ^{U \times 1}}.
    \end{align}
where $\bW(f) \in \mathbb{C}^{M \times U}$ is the baseband combiner, and $\by(f)\in \mathbb{C}^{N_\text{R} \times 1}$ is the signal that impinges on the metamaterial elements at frequency $f$.

\subsection{Spherical-Wave-based Channel Model}\label{sec:Near_Wideband}
Due to the extremely large array aperture, future wireless communications may be implemented in the near-field region. Generally, users are considered to be distributed in the near-field region when the transceiver distance is smaller than the Fraunhofer distance $d_\text{F} \triangleq \frac{2D^2}{\lambda_c}$ but larger than the Fresnel distance $d_{\text{N}} \triangleq \sqrt[3]{\frac{D^4}{8\lambda_c}}$, where $\lambda_c$ is the carrier wavelength and $D$ is the antenna diameter \cite{2010MutualInformation}. When the transceiver distance is larger than $d_\text{F}$, signal transmission is considered to take place in the far-field region. In far-field communications, the spherical wavefront of electromagnetic waves can be well approximated as a plane wavefront for the sake of modeling simplicity. However, such approximation might not hold in the near-field region, and a spherical wave model should be adopted for this case \cite{2009Onspherical}. Also, the spatial-wideband effect resulted from the extremely large-scaled antenna array can be characterized by using the spherical wave model. 
Meanwhile, signals propagating at different frequencies could have different channel gains. Therefore, we will characterize the channel model containing both near-field and dual-wideband effects in the following.

Our work considers a near-field wideband uplink channel where $U$ single-antenna users are located in the near-field region of the HMA-based BS. 
We suppose signals from user $u$ propagate through $P_u+1$ rays where the $0$th ray corresponds to the line-of-sight (LoS) path and $p=1, 2, \ldots, P_u$ rays correspond to non-line-of-sight (NLoS) paths. As is shown in \figref{fig:cartesian_model}, we adopt a Cartesian coordinate system and set the coordinates of the center of the HMA-based array as $(0,0,0)$. We denote coordinates of the $l$th element on the $m$th microstrip as the three-dimensional point $\bp_{m,l}$ and the $p$th scatterer of user $u$ as $\bp_{u,p}$. The total time delay from the $p$th scatterer of user $u$ to the $l$th element on the $m$th microstrip is represented as
\begin{align}\label{eq:time_delay}
    \tau_{u,p,m,l}= \frac{||\bp_{u,p}-\bp_{m,l}||_{\text{2}}}{c},
 \end{align}
where $c = 3 \times 10^8 $ m/s denotes the speed of light.
Let $x_u(t)$ be the baseband signal from user $u$ at time index $t$. Then, the baseband signal observed by the $l$th element on the $m$th microstrip from user $u$ at time index $t$ is formulated in the noiseless case as \cite{2021ChannelWideband}
\begin{align}
    y_{u,m,l}(t) = \sum_{p = 0}^{P_u} \sqrt{\varepsilon_{m,l,u,p}} A_{m,l,u,p}(t) x_u(t- \tau_{u,p,m,l})e^{-\jmath 2 \pi f_c \tau_{u,p,m,l}},
\end{align}
where $f_c$ is the carrier frequency, $\varepsilon_{m,l,u,p}$ and $A_{m,l,u,p}(t)$ are the large-scale fading factor and the channel gain coefficient between the $l$th element on the $m$th microstrip and the $p$th scatterer of user $u$, respectively. We assume that large-scale fading varies slowly with time, so $\varepsilon_{m,l,u,p}$ is independent of $t$ \cite{2013Coordinated}.
Thereafter, the channel between user $u$ and the $l$th element on the $m$th microstrip can be modeled as
\begin{align}\label{eq:g}
    g_{u,m,l}(t) = \sum_{p = 0}^{P_u} \sqrt{\varepsilon_{m,l,u,p}} A_{m,l,u,p}(t) e^{-\jmath 2 \pi f_c \tau_{u,p,m,l}} \delta(t- \tau_{u,p,m,l}).
 \end{align}

With the continuous time Fourier transform (CTFT) of
\eqref{eq:g}, the channel frequency response between user $u$ and the $l$th element on the $m$th microstrip is
\begin{align}\label{eq:gf}
    g_{u,m,l}(f) &= \int_{-\infty}^{+\infty}g_{u,m,l}(t) e^{-\jmath 2 \pi f t} dt
    \ = \sum_{p = 0}^{P_u} \sqrt{\varepsilon_{m,l,u,p}} A_{m,l,u,p}(f)  e^{-\jmath 2 \pi (f_c+f) \frac{||\bp_{u,p}-\bp_{m,l}||_{\text{2}}}{c}}.
 \end{align}
By considering the propagation loss due to signal reflection, we denote the channel gain coefficient model as \cite{2022NearField}, \cite{2021ChannelWideband}
\begin{align}\label{eq:coefficient}
    A_{m,l,u,p}(f)= |\Gamma_{u,p}(f)|\sqrt{F(\Theta_{m,l,u,p})}\frac{c}{4\pi (f + f_c)||\bp_{u,p}-\bp_{m,l}||_{\text{2}}}.
 \end{align}
In \eqref{eq:coefficient}, $\Theta_{m,l,u,p} = (\theta_{m,l,u,p},\phi_{m,l,u,p})$ is the elevation-azimuth pair of the signal of user $u$ from the $p$th scatterer to the $l$th element on the $m$th microstrip, and $F(\Theta_{m,l,u,p})$ is the radiation profile of the corresponding metamaterial element modeled as
\begin{align}\label{eq:radiation}
   F(\Theta_{m,l,u,p}) =  \left\{
            \begin{aligned}
                &2(b+1)\cos^b(\theta_{m,l,u,p}),\quad \theta_{m,l,u,p} \in [0,\frac{\pi}{2}] \\
                &0,\quad \quad \quad \quad \quad \quad \quad \quad \quad \quad  \text{otherwise}
             \end{aligned}
        \right.,
 \end{align}
where $b$ determines the boresight gain.\footnote{For example, in the dipole case we have $b = 2$, which yields $F(\Theta_{m,l,u,p}) = 6\cos^2(\theta_{m,l,u,p})$.} Meanwhile, $\Gamma_{u,p}(f)$ is the reflection coefficient for a rough surface of a scatterer in the $p$th path of user $u$ at frequency $f$, denoted as \cite{2021ChannelWideband}, \cite{2007Scattering}
\begin{align}
    \Gamma_{u,p}(f) =  \left\{
            \begin{aligned} & \frac{\cos\phi_{\text{i},u,p}-n_{\text{t}}\cos\phi_{\text{t},u,p}}{\cos\phi_{\text{i},u,p}+n_{\text{t}}\cos\phi_{\text{t},u,p}}
    e^{-\left( \frac{8\pi^2(f_c+f)^2\sigma_{\text{rough}}^2\cos^2 \phi_{\text{i},u,p}}{c^2} \right)}, p = 1,2,\ldots, P\\
    & 1, \quad \quad \quad \quad \quad \quad \quad \quad \quad \quad \quad  \; \quad \quad \quad \quad \quad \quad \quad \ p = 0
     \end{aligned}
        \right.,
 \end{align}
where $n_{\text{t}}$ is the refractile index, $\sigma_{\text{rough}}$ is the roughness coefficient of the reflecting surface, $\phi_{\text{i},u,p}$ is the incidence and reflection angle in the $p$th path of user $u$, and $\phi_{\text{t},u,p} =  \arcsin( n_{\text{t}}^{-1}\sin\phi_{\text{i},u,p}) $ is the refraction angle in the $p$th path of user $u$.
We denote the channel between the BS and user $u$ into a vector as $\bg_u(f) = {[g_{u,1,1}(f), g_{u,1,2}(f),}$ ${\ldots, g_{u,M,L}(f)]}^H$, and reformulate \eqref{eq:gf} as
\begin{align}\label{eq:gu}
    \bg_u(f) = \sum_{p = 0}^{P_u} \ba_{u,p}(f) \odot \bb_{u,p}(f),
 \end{align}
where
\begin{align}
 \ba_{u,p}(f) = & \left[\sqrt{\varepsilon_{1,1,u,p}} A_{1,1,u,p}(f), \sqrt{\varepsilon_{1,2,u,p}} A_{1,2,u,p}(f), \ldots, \sqrt{\varepsilon_{M,L,u,p}} A_{M,L,u,p}(f)\right],
\end{align}
and
\begin{align}
 \bb_{u,p}(f) = & \left[e^{-\jmath 2 \pi (f_c+f) \frac{||\bp_{u,p}-\bp_{1,1}||_{\text{2}}}{c}}, e^{-\jmath 2 \pi (f_c+f) \frac{||\bp_{u,p}-\bp_{1,2}||_{\text{2}}}{c}}, \ldots, e^{-\jmath 2 \pi (f_c+f) \frac{||\bp_{u,p}-\bp_{M,L}||_{\text{2}}}{c}}\right],\label{eq:array_ response_vector}
 \end{align}
are the channel gains and array response vectors of the BS, respectively. Note that the far-field channel can be derived as a special case of \eqref{eq:gu} when the
transceiver distance $||\bp_{u,p}-\bp_{m,l}||_{\text{2}}$ is sufficiently large.

\subsection{Problem Formulation}\label{sec:Problem_ Formulation}
We consider a total number of $S$ subcarriers over the bandwidth $B$ and denote the $s$th subcarrier frequency as $ f_s = \left(s - \frac{S+1}{2} \right) \frac{B}{S}, s \in \cS \triangleq \{0,1,\ldots, S-1\}$. By defining $\bg_{u,s} = \bg_u[f_s]$, the channel output signal at the $s$th subcarrier frequency can be expressed as
\begin{align}\label{eq:ys}
    \by_s= \sum_{u=1}^{U}{\bg_{u,s} x_{u,s}} + \bn_s = \bG_s \bx_s + \bn_s \in \mathbb{C}^{N_\text{R} \times 1}.
 \end{align}
In \eqref{eq:ys}, $\bG_s \triangleq [\bg_{1,s}, \bg_{2,s}, \ldots, \bg_{U,s}] \in \mathbb{C}^{N_\text{R} \times U}$, $\bx_s \triangleq [x_{1,s}, x_{2,s}, \ldots, x_{U,s}]^T \in \mathbb{C}^{U \times 1}$ where $x_{u,s}$ denotes the transmitted signal of user $u$ at the $s$th subcarrier frequency, and $\bn_s \in {\mathbb{C}^{N_\text{R}\times 1}}$ denotes the independently distributed noise with covariance $\sigma^2_s \bI_{N_\text{R}}$. Combined with \eqref{eq:zf}, the receiving signal $\bz_s$ at the $s$th subcarrier frequency can be formulated as
    \begin{align}\label{eq:received_signal}
    \bz_s = \bW_s^H \bQ \bH_s\bG_s\bx_s + \bW_s^H \bQ \bH_s\bn_s \in {\mathbb{C} ^{U \times 1}}.
    \end{align}

We consider the case where power is uniformly distributed, i.e., $\expect{\bx_s \bx_s^H} = P_{\text{t}}\bI_U$. Additionally, the knowledge of the channel $\bG_s$, $\forall s$, is assumed to be fully known, which can be obtained using similar methods in \cite{2019EstimationVlachos} and \cite{2021CuiNear}.
According to the signal transceiver model in \eqref{eq:received_signal}, the achievable sum rate can be written as
\begin{align}\label{eq:sum rate}
  R_{\text{S}} =  \sum_{s=0}^{S-1}\Delta_B {\log_2 \left| \bI_U + \frac{P_{\text{t}}}{\Delta_B \sigma^2_s} \bW_s^H \bQ \bH_s \bG_s \bG_s^H \bH_s^H \bQ^H \bW_s ( \bW_s^H \bQ \bH_s \bH_s^H \bQ^H \bW_s)^{-1} \right|},
\end{align}
where $\Delta_B$ is the subcarrier separation estimated as $B/S$.

In this paper, we aim to design the baseband combiner $\bW_s$, $\forall s$, and weight matrix $\bQ$ to maximize the sum rate of HMA-based XL-MIMO uplink communications. The corresponding problem can be modeled as
     \begin{subequations}\label{problem_model}
     \begin{align}
       \cP_1:\quad \underset{\bQ,\bW_s,\forall s \in \cS}{ \mathop{\max }}\quad &  \sum_{s=0}^{S-1} {\log_2 \left| \bI_U + \frac{P_{\text{t}}}{\Delta_B \sigma^2_s} \bW_s^H \bQ \bH_s \bG_s \bG_s^H \bH_s^H \bQ^H \bW_s ( \bW_s^H \bQ \bH_s \bH_s^H \bQ^H \bW_s)^{-1} \right|},  \label{problem_model_a}\\
       \mathrm{s.t.} \quad & \left( \mathbf{Q} \right)_{m_1,(m_2-1)L+l}=
           \left\{
            \begin{aligned}
                & q_{m_1,l},\quad m_1=m_2 \\
                & 0,\quad  \quad \ m_1\ne m_2 \\
             \end{aligned}
            \right. , \label{Q_stucture}\\
          &  q_{m_1,l} \in \cQ, \; \forall m_1,l \label{q_stucture},
      \end{align}
      \end{subequations}
where $m_1 = 1, 2, \ldots, M$, $m_2 = 1, 2, \ldots, M$, and $l = 1, 2, \ldots, L$. Since the coefficient $\Delta_B$ before the sum log function is a constant, it has been omitted in problem $\cP_1$ without loss of generality.
Problem $\cP_1$ is a challenging problem that cannot be addressed by conventional beam combining algorithms for the following reasons. Firstly, since the baseband combiner $\bW_s$,$\forall s \in \cS$, and the weight matrix $\bQ$ are nonlinearly coupled, optimizing the two matrices simultaneously is difficult. Secondly, $\bQ$ is a block matrix constrained by \eqref{Q_stucture}, which is difficult to handle.  Finally, the feasible set of weights \eqref{q_stucture} is non-convex, thus further complicating the problem. In the next section, we will propose algorithms to address these difficulties.

\section{Proposed Algorithm for HMA-Assisted Beam Combining}\label{sec:mathematical model}
In this section, we propose an algorithm for the HMA-based XL-MIMO uplink beam combining, which can effectively address the nonlinear coupling of variables and non-convexity of constraints. Specifically, to decouple variables $\bQ$ and $\bW_s$, $s \in \cS$, we adopt an AO method to optimize them in an iterative manner. Besides, since the sum-rate maximization objective is complex, we adopt the MMSE approach to transform it to a relatively simple equivalent objective. Meanwhile, we adopt the matrix vectorization and MM method to tackle the block structure in \eqref{Q_stucture} and non-convex constraint in \eqref{q_stucture}. In the following, we will illustrate these methods in detail.

\subsection{Equivalent Matrix-Weighted MMSE}\label{sec:MWMMSE}
Problem $\cP_1$ is a complex sum-rate maximization problem, which can be equivalent to the matrix-weighted MMSE one as \cite{2011IterativelyWMMSE}
     \begin{subequations}\label{P_2}
     \begin{align}
       \cP_2:\quad \underset{\bQ, \bW_s,\bM_s,\forall s \in \cS}{ \mathop{\min }}\quad &  \sum_{s=0}^{S-1} {\tr{ \bM_s \bE_s(\bQ, \bW_s)} - \log_2|\bM_s| } ,\label{eq:P21}\\
       \mathrm{s.t.} \quad & \left( \bQ \right)_{m_1,(m_2-1)L+l}=
           \left\{
            \begin{aligned}
                & q_{m_1,l},\quad m_1=m_2 \\
                & 0,\quad \  \quad m_1\ne m_2 \\
             \end{aligned}
            \right. ,\\
          &  q_{m_1,l} \in \cQ, \; \forall m_1,l.
      \end{align}
      \end{subequations}
In $\cP_2$, $\bM_s$ is a weight matrix that satisfies $\bM_s \succeq \bzero$ and $\bE_s(\bQ, \bW_s)$ is the mean-square error matrix given by
\begin{align}\label{eq:E}
    \bE_s(\bQ, \bW_s)  &= \expect{(\bz_s - \bx_s)(\bz_s - \bx_s)^H} \ntb
    &= \expect{\left(\bW_s^H \bQ \bH_s (\bG_s \bx_s + \bn_s) -\bx_s \right) \left(\bW_s^H \bQ \bH_s (\bG_s \bx_s + \bn_s) -\bx_s\right)^H}\ntb
    &= P_{\text{t}} (\bW_s^H \bQ \bH_s \bG_s - \bI_U)(\bW_s^H \bQ \bH_s \bG_s  - \bI_U)^H + \Delta_B \sigma^2_s \bW_s^H \bQ \bH_s \bH_s^H \bQ^H \bW_s.
\end{align}
Note that $\cP_2$ is easier to handle than $\cP_1$ since the objective function \eqref{eq:P21} is convex with respect to $\bW_s$ or $\bM_s$. As the variables $\bQ$, $\bW_s$, and $\bM_s$, $\forall s \in \cS$, are coupled with each other, we adopt an AO method to optimize each variable by fixing the other two variables. When $\bQ$ and $\bW_s$, $\forall s \in \cS$, are given, the optimal $\bM_s$, $\forall s \in \cS$, can be obtained by the first-order optimality condition of the Lagrangian function \cite{2011IterativelyWMMSE}. In particular, the optimal $\bM_s$ is given in a closed-form
\begin{align}\label{eq:Msopt}
    \bM_s^{\text{opt}} = \bE_s^{-1}(\bQ, \bW_s).
\end{align}
Similarly, with given $\bM_s$, $\forall s \in \cS$, and $\bQ$, the optimal $\bW_s$, $\forall s \in \cS$, is formulated as
\begin{align}\label{eq:Wsopt}
    \bW_s^{\text{opt}} = (P_{\text{t}} \bQ \bH_s \bG_s\bG_s^H \bH_s^H \bQ^H + \Delta_B \sigma^2_s \bQ \bH_s \bH_s^H \bQ^H)^{-1}\bQ \bH_s \bG_s.
\end{align}

When optimizing $\bQ$ with given $\bM_s$ and $\bW_s$, $\forall s \in \cS$, we plug \eqref{eq:E} into problem $\cP_2$, and obtain the optimization problem as
 \begin{subequations}
 \begin{align}
   \cP_3:\quad \underset{\bQ}{\mathop{\min }}\quad &  \sum_{s=0}^{S-1}  \tr{P_{\text{t}} \bM_s \bW_s^H \bQ \bH_s \bG_s \bG_s^H \bH_s^H \bQ^H \bW_s} - \tr{P_{\text{t}} \bM_s \bW_s^H \bQ \bH_s \bG_s } \ntb
   & - \tr{P_{\text{t}} \bM_s \bG_s^H \bH_s^H \bQ^H \bW_s} + \tr{P_{\text{t}} \bM_s} + \tr{\Delta_B \sigma^2_s \bM_s \bW_s^H \bQ \bH_s \bH_s^H \bQ^H \bW_s}\ntb
   & - \log_2|\bM_s| \label{eq:objective_p3},\\
  \mathrm{s.t.} \quad & \left( \bQ \right)_{m_1,(m_2-1)L+l}=
   \left\{
    \begin{aligned}
        & q_{s,m_1,l},\quad m_1=m_2 \\
        & 0,\quad \quad  \quad m_1\ne m_2 \\
     \end{aligned}
    \right. ,\label{eq:Q_constraint_p3} \\
  &  q_{m_1,l} \in \cQ, \; \forall m_1,l. \label{eq:Lorentzian}
  \end{align}
  \end{subequations}
Problem $\cP_3$ is still a challenging task since the feasible $\bQ$ should satisfy the block structure \eqref{eq:Q_constraint_p3} and some specific forms \eqref{eq:Lorentzian}. This motivates us to transform the objective function by the matrix vectorization method to eliminate the block structure and adopt the MM method to find the feasible weights. We present the details in the following.

\subsection{Matrix Vectorization}\label{sec:Matrix_ Vectorization}
For convenience, we define $\bB_s = \bW_s\bM_s \bW_s^H$, $\bA_s = \bH_s \bG_s \bG_s^H \bH_s^H$, $\bC_s = \bH_s\bG_s \bM_s \bW_s^H$, and $\bD_s = \bH_s \bH_s^H$. Then, the objective function in $\cP_3$ can be written as
 \begin{align} \label{eq:fQ}
   \text{f}(\bq) = \sum_{s=0}^{S-1}  P_{\text{t}} \tr{\bQ^H \bB_s \bQ \bA_s} - P_{\text{t}} \tr{ \bQ \bC_s}  - P_{\text{t}} \tr{ \bQ^H \bC_s^H} + \Delta_B \sigma^2_s \tr{\bQ^H \bB_s \bQ \bD_s} ,
  \end{align}
where we omit the terms $\tr{P_{\text{t}} \bM_s}$ and $\log_2|\bM_s|$ that are independent of $\bQ$. Note that since $\bM_s$ is a Hermitian matrix according to \eqref{eq:Msopt}, i.e., $\bM_s = \bM_s^H$, we have $\bC_s^H = \bW_s \bM_s \bG_s^H \bH_s^H$. To remove the block structure in \eqref{eq:Q_constraint_p3}, we define $\bq = [q_{1,1}, q_{1,2}, \ldots, q_{m,(m-1)L+l}, \ldots, q_{M,ML}]^T$ where $q_{m,(m-1)L+l}$ is the $\left(m,\left(m-1\right)L+l \right)$th entry of $\bQ$. We also define ${\bc_s = [(\bC_s)_{1,1}, (\bC_s)_{2,1}, \ldots, }$ ${ (\bC_s)_{(m-1)L+l,m}, \ldots(\bC_s)_{ML,M}]^T}$. Additionally, we have
    \begin{subequations}
    \begin{align}
        &\tr{ \bQ \bC_s} = \bq^T \bc_s,  \quad \tr{\bQ^H \bC_s^H} = \bc_s^H \bq^{*}, \label{eq:tr2} \\
        &\tr{ \bQ^H \bB_s \bQ \bA_s } = \bq^H \left( \bB_s \otimes \bN_L\right) \odot \bA_s^T \bq, \label{eq:tr1}\\
         &\tr{ \bQ^H \bB_s \bQ \bD_s} = \bq^H \left( \tilde{\bB}_s \otimes \bI_L \right) \odot \bD_s \bq, \label{eq:tr3}
    \end{align}
    \end{subequations}
where $\bN_L$ is a $L \times L$ all-one matrix and $\tilde{\bB}_s$ is a diagonal matrix that contains the principal diagonal elements of matrix $\bB_s$, i.e., $(\tilde{\bB}_s)_{m,m} = (\bB_s)_{m,m}$. The proof of Eqs. \eqref{eq:tr2}, \eqref{eq:tr1}, and \eqref{eq:tr3} are provided in Appendix \ref{sec:Appendix_A}.

By plugging Eqs. \eqref{eq:tr2}, \eqref{eq:tr1}, and \eqref{eq:tr3} into \eqref{eq:fQ}, we have
 \begin{align}\label{eq:fq}
  \text{f}(\bq)= \sum_{s=0}^{S-1}  \bq^H \left( P_{\text{t}} (\bB_s \otimes \bN_L) \odot \bA_s^T + \Delta_B \sigma^2_s \left(\tilde{\bB}_s  \otimes \bI_L \right)\odot \bD_s \right)\bq  - 2P_{\text{t}}\operatorname{Re}\{\bq^H \bc_s^{*}\}.
  \end{align}
We define $\bS = \sum_{s=0}^{S-1}P_{\text{t}}(\bB_s \otimes \bN_L) \odot \bA_s^T +\Delta_B \sigma^2_s \left( \tilde{\bB}_s  \otimes \bI_L \right)\odot \bD_s $ and $\bc = \sum_{s=0}^{S-1}P_{\text{t}}\bc_s$. Then, problem $\cP_3$ is transformed into
 \begin{subequations}\label{eq:P4}
 \begin{align}
   \cP_4: \quad
    \underset{\bq}{\mathop{\min }}\quad &  \bq^H \bS \bq  - 2 \operatorname{Re}\{\bq^H \bc^{*}\}, \label{eq:objective_p4}\\
      \mathrm{s.t.} \quad  &  q_{m_1,l} \in \cQ, \; \forall m_1,l.
  \end{align}
  \end{subequations}
As mentioned in Subsection \ref{sec:HMA}, there are three typical models for the frequency response of metamaterial elements, including the amplitude-only weight, binary amplitude weight, and Lorentzian-constrained phase. In addition to these cases, we also consider an ideal case where the feasible set of HMA weights is the complex plane. In the following, we will discuss the optimization of HMA weights for these four cases, respectively.

\subsection{Optimization of HMA Weights}\label{sec:MM}

\subsubsection{Unconstrained Weight Case}
For the unconstrained weight case, i.e., the ideal case, the problem in \eqref{eq:P4} can be written as
 \begin{subequations}
 \begin{align}
   \cP_5: \quad
    \underset{\bq}{\mathop{\min }}\quad & \bq^H \bS \bq  - 2 \operatorname{Re}\{ \bc^{T}\bq\},\\
      \mathrm{s.t.} \quad  &  q_{m_1,l} \in \C, \; \forall m_1,l.
  \end{align}
  \end{subequations}
Problem $\cP_5$ is concave and can be solved by conventional convex optimization methods \cite{2004ConvexBoyd}.

\subsubsection{Amplitude-Only Weight Case}
When the electrical circuit is near resonance, the tuning modality of amplitude-only weight is possible for each metamaterial element \cite{2017Analysis}. For amplitude-only weights, all entries of $\bq$ have real values within $[a, b]$, $b \textgreater a \textgreater 0$. Then, problem $\cP_4$ can be concretized as
 \begin{subequations}
 \begin{align}
   \cP_6: \quad
    \underset{\bq}{\mathop{\min }}\quad & \bq^T \bS \bq  - 2 \operatorname{Re}\{ \bc^{T}\}\bq,\\
      \mathrm{s.t.} \quad  &  q_{m_1,l} \in \left[a, b \right], \; b \textgreater a \textgreater 0, \; \forall m_1,l.
  \end{align}
  \end{subequations}
Problem $\cP_6$ yields a maximization of a concave function with linear constraints. Therefore, it can also be efficiently solved by classical convex optimization methods \cite{2004ConvexBoyd}.
\subsubsection{Binary Amplitude Weight Case}
One of the more easily tuning methods
is to switch each metamaterial element between ``on" and ``off" states, which results in only two feasible amplitudes for each element \cite{2017DesignYurduseven}. For binary amplitude weights, all entries of $\bq$ belong to $c \cdot \{0, 1\}$, $c \textgreater 0$. Therefore, $P_4$ can be specified as
  \begin{subequations}
 \begin{align}
   \cP_7: \quad
    \underset{\bq}{\mathop{\min }}\quad &  \bq^T \bS \bq  - 2 \operatorname{Re}\{ \bc^{T}\}\bq, \\
      \mathrm{s.t.} \quad  &  q_{m_1,l} \in c \cdot \{0, 1\}, \; c \textgreater 0, \; \forall m_1,l.
  \end{align}
  \end{subequations}
There are many ways to arrive at on/off elements that can potentially provide a directed beam solution, including the exhaustive search method.

\subsubsection{Lorentzian-Constrained Phase Case}
It is well-known in beamforming that control over the phase generally provides a better beam than what is accomplished by only amplitudes \cite{2017Analysis}. When considering the Lorentzian-constrained phases, we can reformulate $P_4$ as
  \begin{subequations}
 \begin{align}
   \cP_8: \quad
    \underset{\bq}{\mathop{\min }}\quad &  \bq^H \bS \bq  - 2 \operatorname{Re}\{ \bq^H \bc^{*}\}, \\
      \mathrm{s.t.} \quad  &  q_{m_1,l} \in \{ \frac{\jmath+e^{\jmath \phi}}{2} : \phi \in [0, 2\pi] \}, \; \forall m_1,l.
  \end{align}
  \end{subequations}
We represent $\bq = \frac{1}{2}\left( \jmath \bm{1}_{N_{\text{R}}} +\bp\right)$ where $\bm{1}_{N_{\text{R}}}$ is an all-one vector, hence the entries of $\bp \in \C^{N_{\text{R}}}$ are all unit modulus, i.e., $p_{(m-1)L+l} \in \{ e^{\jmath \phi}, \phi \in [0,2\pi] \}, \; \forall m = 1,2, \ldots, M, \forall l = 1,2, \ldots, L$. By plugging $\bq = \frac{1}{2}\left( \jmath \bm{1}_{N_{\text{R}}} +\bp\right)$ into \eqref{eq:objective_p4}, we have
 \begin{align}
  \text{f}(\bp)= \frac{1}{4}\bp^H \bS \bp  + \operatorname{Re}\{ \frac{\jmath}{2}\bp^H \bS \bm{1}_{N_{\text{R}}}- \bp^H \bc^{*}\} +  \operatorname{Re}\{\jmath \bm{1}_{N_{\text{R}}}^T \bc^{*}\}
   + \frac{1}{4}\bm{1}_{N_{\text{R}}}^T \bS \bm{1}_{N_{\text{R}}}. \label{eq:fp}
  \end{align}
Thus, the optimization variable of problem $\cP_8$ is simplified from Lorentzian-constrained phase $\bq$ to the unit modulus phase $\bp$.

In \eqref{eq:fp}, $\text{f}(\bp)$ is a quadratic function due to the first term $\frac{1}{4}\bp^H \bS \bp$. We resort to the MM method, a sequential convex optimization approach, to obtain a suboptimal solution to problem $\cP_8$. Since a tractable surrogate subproblem leads to high effectiveness, we aim to find a tractable surrogate function to approximate $\text{f}(\bp)$ in the following.

\begin{lemma}\label{theorem:1}
Define $\bT$ as a $N_\text{R} \times N_\text{R}$ Hermitian matrix such that $\bT \succeq \bS $. For any given $\bp^{(\ell)}$ at the $(\ell-1)$th iteration, we have
    \begin{align}
    \bp^H\bS\bp \leq \bp^H \bT \bp - 2\operatorname{Re}\{\bp^H (\bT-\bS)\bp^{(\ell)}\} + (\bp^{(\ell)})^H (\bT-\bS)\bp^{(\ell)}.
    \end{align}
\end{lemma}
The proof of \emph{\lmref{theorem:1}} can be found in \cite[Lemma III.2]{2020Hybrid}, thus is omitted here.

Inspired by the inequality in \emph{\lmref{theorem:1}}, we choose $\bT = \lambda_{\max}\bI$, where $\lambda_{\max}$ is the maximum eigenvalue of $\bS$ \cite{2021Reconfigurable,2021RISTradeoff}. Then, we get
\begin{align}\label{eq:surrogate_function}
    \text{f}(\bp)  \leq \tilde{\text{f}}(\bp |\bp^{(\ell)}) = & \frac{\lambda_{\max}}{4} \bp^H \bp - \frac{1}{2}\operatorname{Re}\{\bp^H (\lambda_{\max}\bI-\bS)\bp^{(\ell)}\} + \frac{1}{4}(\bp^{(\ell)})^H (\lambda_{\max}\bI-\bS)\bp^{(\ell)}
     \ntb
     & \ + \operatorname{Re}\{ \frac{\jmath}{2}\bp^H \bS \bm{1}_{N_{\text{R}}}- \bp^H \bc^{*}\} + \operatorname{Re}\{\jmath \bm{1}_{N_{\text{R}}}^T \bc^{*}\}
   + \frac{1}{4}\bm{1}_{N_{\text{R}}}^T \bS \bm{1}_{N_{\text{R}}}.
\end{align}
The function $\tilde{\text{f}}(\bp |\bp^{(\ell)})$ is a valid majorizer of $\text{f}(\bp)$ \cite{2017MM}. By adopting the function $\tilde{\text{f}}(\bp |\bp^{(\ell)})$ and MM framework, we obtain a surrogate problem of $\cP_8$ as
    \begin{subequations}
    \begin{align}
     \cP_9^{(\ell)}: \quad \underset{\bp}{\min} &\quad \tilde{\text{f}}(\bp |\bp^{(\ell)}), \\
            \mathrm{s.t.}&\quad  p_{(m-1)L+l} \in \{  e^{\jmath \phi}, \phi \in [0,2\pi] \}, \; \forall m,l.
    \end{align}
    \end{subequations}
Note that the first term of $\tilde{\text{f}}(\bp |\bp^{(\ell)})$ in \eqref{eq:surrogate_function} is a constant, i.e., $\bp^H \bp = N_\text{R}$. Also, the corresponding third, fifth, and sixth terms are independent of $\bp$. Therefore, problem $\cP_9^{(\ell)}$ can be simplified as
    \begin{subequations}
    \begin{align}
     \cP_{10}^{(\ell)}: \quad \underset{\bp}{\max} &\quad \operatorname{Re}\left\{\bp^H \left[ (\lambda_{\max}\bI_{N_\text{R}}-\bS)\bp^{(\ell)} +2\bc^{*} - \jmath \bS \bm{1}_{N_{\text{R}}} \right] \right \}, \\
     \mathrm{s.t.}&\quad  p_{(m-1)L+l} \in \{  e^{\jmath \phi}, \phi \in [0,2\pi] \}, \; \forall m,l.
    \end{align}
    \end{subequations}
We define $\ba^{(\ell)} = (\lambda_{\max}\bI_{N_\text{R}}-\bS)\bp^{(\ell)} +2\bc^{*} - \jmath \bS \bm{1}_{N_{\text{R}}}$, and represent its $n$th element as $(\ba^{(\ell)})_n = |a^{(\ell)}_n | e^{\jmath \arg a^{(\ell)}_n}$. Also, we represent the $n$th element of $\bp^H$ as $p_n^{*} = e^{-\jmath \phi_n}$. It is straightforward to find a closed-form solution to problem $\cP_{10}^{(\ell)}$ as
\begin{align}
  \phi_n =  \arg a^{(\ell)}_n, \; \forall n.
\end{align}
Therefore, a suboptimal solution to $\cP_8$ can be obtained by calculating $\ba^{(\ell)} = (\lambda_{\max}\bI_{N_\text{R}}-\bS)\bp^{(\ell)} +2\bc^{*} - \jmath \bS \bm{1}_{N_{\text{R}}}$ and $p_n^{(\ell+1)} = e^{\jmath \arg a^{(\ell)}_n}$, $\forall n$, in an alternating manner, where details are described in \textbf{\alref{alg:MM}}.
\begin{algorithm}[t]
\caption{MM-based Lorentzian-Constrained Phase Optimization} 
 \label{alg:MM}
 \begin{algorithmic}[1]
\Require  The matrix $\bS$, vector $\bc$, and threshold $\epsilon$.
\State Initialize $\bp^{(\ell)}$ with unit modulus and set iteration index $\ell=0$.
\State Calculate $\text{f}(\bp^{(\ell)})$. 
 \Repeat 
    \State  Calculate $\ba^{(\ell)} = (\lambda_{\max}\bI_{N_\text{R}}-\bS)\bp^{(\ell)} +2\bc^{*} - \jmath \bS \bm{1}_{N_{\text{R}}}$.
    \State Calculate the $n$th element of $\bp^{(\ell+1)}$ as $p^{(\ell+1)}_n = e^{\jmath \arg a^{(\ell)}_n}$, $n = 1, 2, \ldots N_\text{R}$.
    \State Calculate $\text{f}(\bp^{(\ell+1)})$.
    \State $\ell \leftarrow \ell+1$.
 \Until{$\left| \text{f}(\bp^{(\ell)})- \text{f}(\bp^{(\ell-1)}) \right|\leq \epsilon $}
\Ensure The HMA weight vector $\bq = \frac{\jmath + \bp^{(\ell)}}{2}$ and the weight matrix $\bQ$.
\end{algorithmic}
\end{algorithm}

\subsection{Convergence and Complexity Analysis}\label{subsec:Overall_algorithm}
This section presents an algorithmic framework for designing near-field wideband beam combining in HMA-based XL-MIMO uplink systems. In particular, since the sum-rate maximization in problem $\cP_1$ has a complex objective function, we adopt its equivalent objective function to get a simple matrix-weighted MMSE problem. Considering the nonlinear coupling of HMA weights and baseband combiners, we iteratively optimize them based on the AO method. Specifically, the baseband combiners can be obtained in a closed-form solution by \eqref{eq:Wsopt} with arbitrary HMA weights. When baseband combiners are given, HMA weights can be obtained by the matrix vectorization and MM methods, which overcome challenges caused by the block structure and some specific constraints for HMA weights. We present the complete algorithm for the near-field wideband beam combining of HMA-based XL-MIMO uplink systems in \textbf{\alref{alg:overallalg}}.

\begin{algorithm}[t]
\caption{HMA-based Wideband Beam Combining Algorithm} 
 \label{alg:overallalg}
 \begin{algorithmic}[1]
\Require  The channel $\bG_s$, diagonal matrices $\bH_s$, transmit power budget $P_{\text{t}}$, and noise power density $\sigma_s^2$, $\forall s$, as well as the subcarrier separation $\Delta_B$, subcarrier number $S$, and threshold $\epsilon$.

\State Initialize the baseband combiner $\bW_s^{(\ell)}$,  $\forall s \in \cS$, HMAs' weight matrix $\bQ^{(\ell)}$, and auxiliary weight matrix $\bM_s^{(\ell)}$, $\forall s \in \cS$. Set iteration index as $\ell=0$.

\State Calculate $R_{\text{S}}^{(\ell)}$. 
 \Repeat 
  \State  Calculate $\bW_s^{(\ell+1)}$, $\forall s \in \cS$, by using \eqref{eq:Wsopt} and $\bQ^{(\ell)}$.

    \State  Calculate $\bE_s^{(\ell+1)}$ by using \eqref{eq:E}, $\bQ^{(\ell)}$ and $\bW_s^{(\ell+1)}$, $\forall s \in \cS$.

    \State Calculate $\bM_s^{(\ell+1)} = (\bE_s^{(\ell+1)})^{-1}$, $\forall s \in \cS$.

    \State Calculate $\bQ^{(\ell+1)}$ by solving $\cP_6$, $\cP_7$, or $\cP_8$ with $\bM_s^{(\ell+1)}$ and $\bW_s^{(\ell+1)}$, $\forall s \in \cS$.
%

    \State Calculate $R_{\text{S}}^{(\ell+1)}$.

    \State $\ell \leftarrow \ell+1$.

 \Until{$\left| R_{\text{S}}^{(\ell)}- R_{\text{S}}^{(\ell-1)} \right|\leq \epsilon $.}
\Ensure The baseband combiners $\bW_s$, $\forall s \in \cS$, and HMA weight matrix $\bQ$.
\end{algorithmic}
\end{algorithm}

%
%
Since \textbf{\alref{alg:MM}} and \textbf{\alref{alg:overallalg}} are both based on the AO method, it is essential to prove the convergence of the proposed algorithm. For \textbf{\alref{alg:MM}}, $\text{f}(\bp^{(\ell)})$, $\forall \ell$, is a non-increasing sequence since $\text{f}(\bp^{(\ell+1)}) \leq \tilde{\text{f}}(\bp^{(\ell+1)} | \bp^{(\ell)}) \leq \tilde{\text{f}}(\bp^{(\ell)} | \bp^{(\ell)}) = \text{f}(\bp^{(\ell)})$ \cite{2017MM}. Additionally, any limit point $(\bM_s, \bQ, \bW_s)$, $\forall s  \in S$ in the iterations by \textbf{\alref{alg:overallalg}} is a stationary point of \eqref{eq:P21} \cite[Theorem 3]{2011IterativelyWMMSE}. Based on these facts, both solutions to $\bQ$ and $\bW_s$, $\forall s \in \cS$ will not increase the value of the objective function in $\cP_2$. Hence, the convergence of both \textbf{\alref{alg:MM}} and \textbf{\alref{alg:overallalg}} is guaranteed.

The computational complexity of \textbf{\alref{alg:overallalg}} mainly comes from the matrix products and inversion for calculating $\bW_s$, $\bE_s$, $\bM_s$, $\forall s \in \cS$, and HMA weights. It is easy to obtain the complexities of calculating $\bW_s$, $\bE_s$, and $\bM_s$ at the whole subcarrier frequencies, which are $\cO(SN_{\text{R}}^2M)$, $\cO(SN_{\text{R}}^2U)$, and $\cO(S U^3)$, respectively. For the unconstrained or
amplitude-only weight matrix, its optimization needs to tackle a convex program with $N_{\text{R}}$ variables. The corresponding complexity can be estimated as $\cO(N_{\text{R}}^p)$, where $1 \leq p \leq 4$ for the standard convex program \cite{2021Reconfigurable}. For
binary amplitude weights, the corresponding problem needs to optimize $N_{\text{R}}$ variables, too. Since we adopt the exhaustive search method to obtain binary amplitude weights, the computational complexity can be estimated as $\cO(N_{\text{R}}^3)$.
For Lorentzian-constrained phases, \textbf{\alref{alg:MM}} starts with the eigenvalue decomposition of the $N_{\text{R}} \times N_{\text{R}}$ matrix $\bS$, whose computational complexity can be estimated as $\cO(N_{\text{R}}^3)$. Then, we suppose that the calculation of $\ba$ and $\bp$ achieves convergence after $I_{\text{MM}}$ iterations. The complexity of each iteration mainly comes from calculating $\ba$, which is given by $\cO(N_{\text{R}}^2)$. Therefore, the complexity of calculating Lorentzian-constrained phases is approximated as $\cO( N_{\text{R}}^3 + I_{\text{MM}} N_{\text{R}}^2)$. Suppose that optimizing the baseband combiners, $\bW_s$, $\forall s \in \cS$, and HMAs' weight $\bQ$, requires $I_{\text{AO}}$ iterations to converge in
total. Then, the overall computational complexity of \textbf{\alref{alg:overallalg}} for four kinds of HMA weights can be estimated as $ \cO\left( I_{\text{AO}} ( S N_{\text{R}}^2 M + N_{\text{R}}^p \right)$, $ \cO\left( I_{\text{AO}} ( S N_{\text{R}}^2 M + N_{\text{R}}^p \right)$, $ \cO \left( I_{\text{AO}} (S N_{\text{R}}^2 M +  N_{\text{R}}^3 )\right)$, and $ \cO\left( I_{\text{AO}} ( S N_{\text{R}}^2 M + N_{\text{R}}^3 + I_{\text{MM}} N_{\text{R}}^2 )\right)$, respectively.

\section{Simulation Results}\label{sec:Simulation}
In this section, we evaluate the performance of the proposed algorithms for HMA-based near-field wideband XL-MIMO uplink communications. At the BS, we adopt an $A$ $\times$ $A$ HMA-based array with $A$ denoting the array length and define $D = \sqrt{2}A$ as the array aperture. The spacing between adjacent metamaterial elements is set as $\lambda /5$ on a microstrip, and the spacing between adjacent microstrips is set as $\lambda /2 $. Therefore, there are $M = \lfloor 2A/\lambda \rfloor$ microstrips and $L = \lfloor 5A/\lambda \rfloor$ metamaterial elements per microstrip. In the conventional antenna array, the spacing between adjacent antenna elements is set as $\lambda /2$. We assume that all metamaterial elements have the same frequency selectivity, i.e., $\bH_s = h_s \bI$ \cite{2019Dynamicuplink}. For the parameters of the spherical-wave-based channel, we suppose that signals from the $u$th user propagate through $1$ LoS path and $P_u = 5$ NLoS paths.
We set the carrier frequency as $f_c = 26$ GHz, the number of subcarriers as $S = 12$, and the noise power density as $\sigma^2_s = -174$ dBm/Hz, $\forall s$ \cite{2021ChannelWideband}, \cite{2019Beam}. Additionally, the large scale fading factor $\varepsilon_{m,l,u,p}$ is set as $10 \log (\varepsilon_{m,l,u,p}) = -38 \log(d_{m,l,u,p}) - 34.5 + \eta_{m,l,u,p}$, where $d_{m,l,u,p}$ denotes the transceiver distance and $\eta_{m,l,u,p}$ denotes the log-normal shadow fading with zero mean and standard deviation $8$ dB \cite{2013Coordinated}. The refractile index $n_t$ is set as $2.24 - \jmath0.025$, and the roughness coefficient $\sigma_{\text{rough}}$ is set as $0.088 \cdot 10^{-3}$ m \cite{2021ChannelWideband}.
Besides, we focus on the following feasible sets for HMA weights \cite{2019Dynamicuplink}:
\begin{itemize}
  \item UC:$\ $ unconstrained, i.e., $\cQ= \C$;
  \item AO:$\ $ amplitude-only, i.e., $\cQ= \left[0.001,5  \right]$;
  \item BA:$\ $ binary amplitude, i.e., $\cQ= \{0, 0.1\}$;
  \item LP:$\ $ Lorentzian-constrained phase, i.e., ${\cQ= \{ \frac{\jmath+e^{\jmath \phi}}{2}:\phi \in [0, 2\pi] \}}$.
\end{itemize}

\subsection{Convergence Performance}
\begin{figure}
\centering
\subfloat[]{\centering\includegraphics[width=0.48\textwidth]{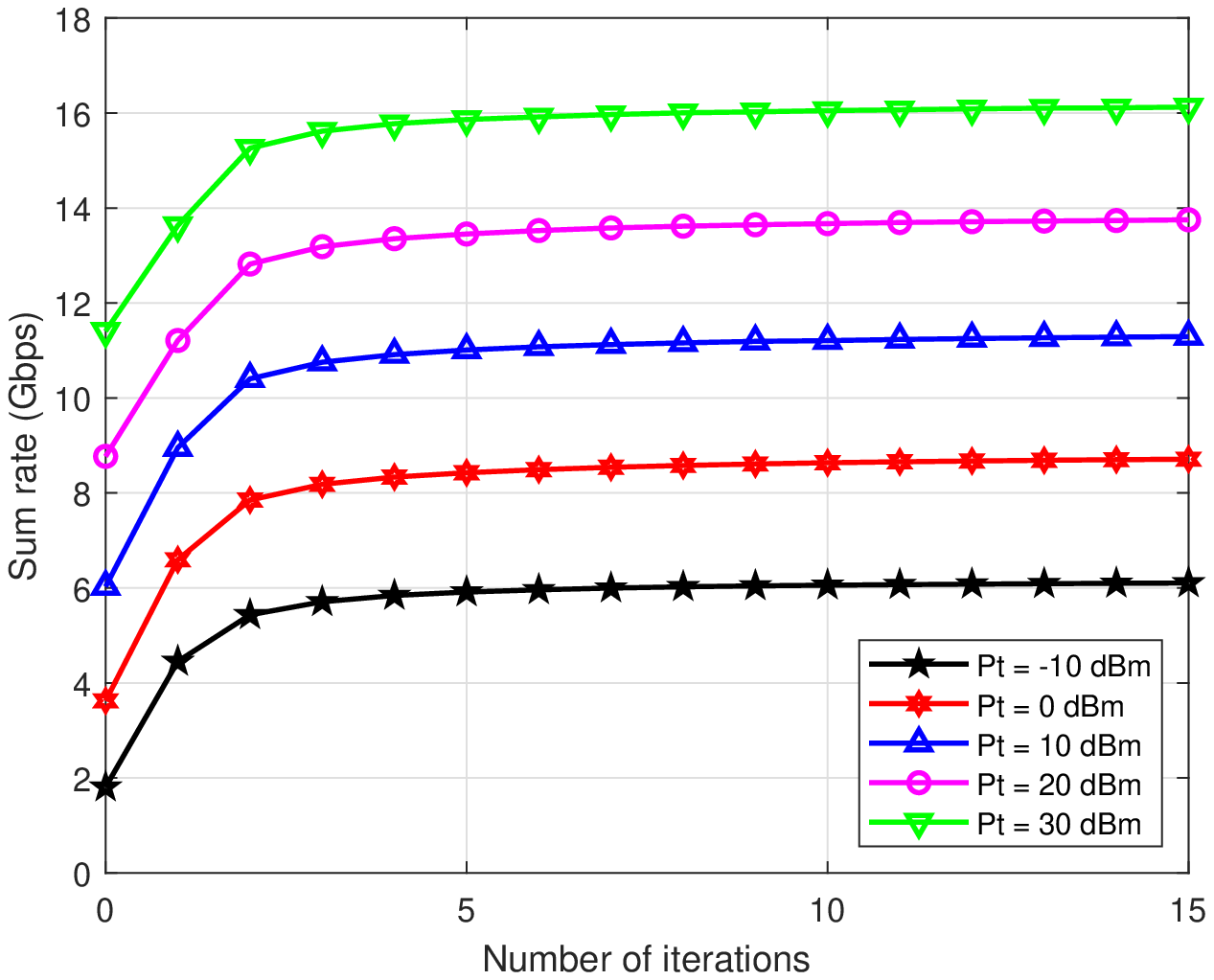}\label{fig:iterationAO}}\hfill
\subfloat[]{\centering\includegraphics[width=0.48\textwidth]{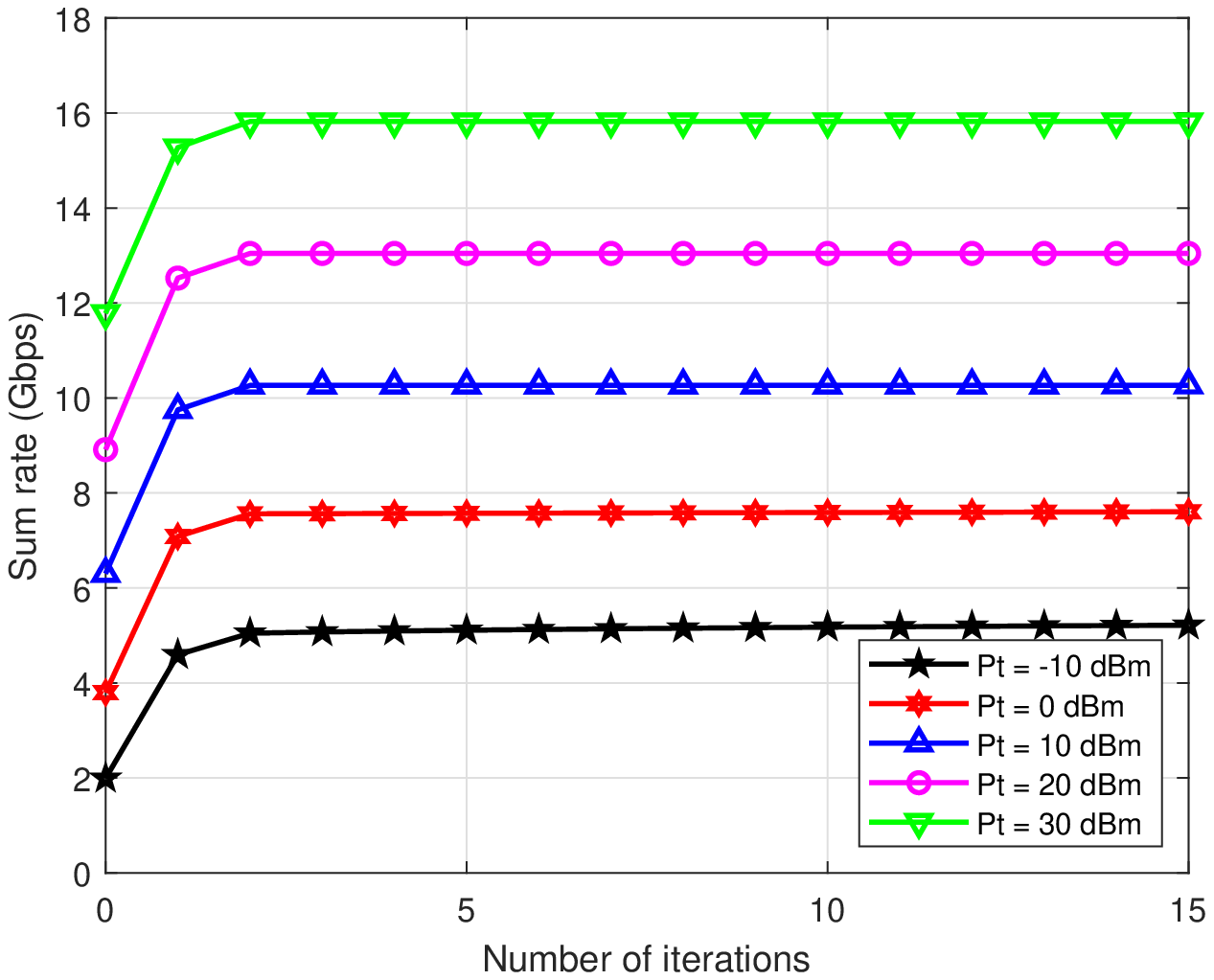}\label{fig:iterationLP}}
\caption{Convergence performance of \textbf{\alref{alg:overallalg}} with different power budgets: (a) amplitude-only weights; (b) Lorentzian-constrained phases.}
\label{fig:Convergence}
\end{figure}

The convergence performance of \textbf{\alref{alg:overallalg}}, i.e., alternatingly designing $\bQ$ and $\bW_s$, $\forall s \in \cS$, budgets is presented in \figref{fig:Convergence} with different transmit power.
Here, we choose amplitude-only and Lorentzian-constrained phase sets for the proposed beam combining design in HMA-assisted XL-MIMO systems. From \figref{fig:Convergence}\subref{fig:iterationAO} and \subref{fig:iterationLP}, we can observe that the proposed algorithms provide a non-decreasing sum-rate sequence for both cases. In addition, the sum rates usually converge after several iterations for different power budgets, which indicates the good convergence performance of the proposed algorithms.

\subsection{Dual-Wideband Effects}

In this subsection, we evaluate the dual-wideband effects on the HMA-based XL-MIMO beam combining.
We suppose that $U = 2$ users are randomly distributed on the $xz$-plane in the near-field region with the angle of arrival $\theta_{u,p} \in [-\pi/4, \pi/4]$. We use ``baseline" to represent the sum-rate performance of the conventional beam combining design that neglects the dual-wideband effects in the HMA-assisted XL-MIMO uplink communications.



\begin{figure*}[t]
    \begin{minipage}{0.49\linewidth}
    \includegraphics[width=8.2cm]{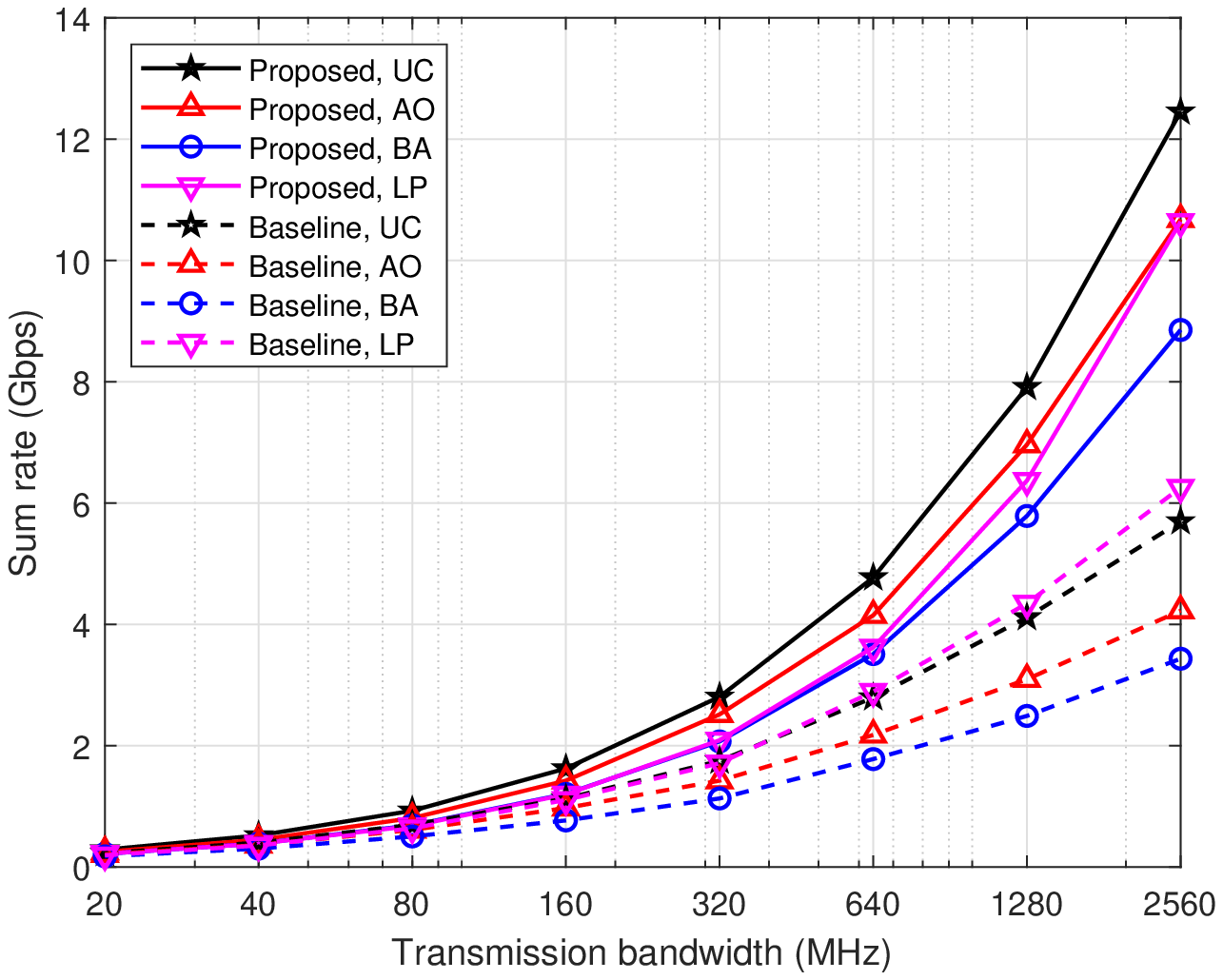}
    \caption{The comparison between the sum-rate performance of the proposed algorithms and baselines for HMA-assisted XL-MIMO communications versus the transmission bandwidth.}\label{fig:wideband1}
    \end{minipage}
    \hfill
    \begin{minipage}{0.49\linewidth}
    \includegraphics[width=8.2cm]{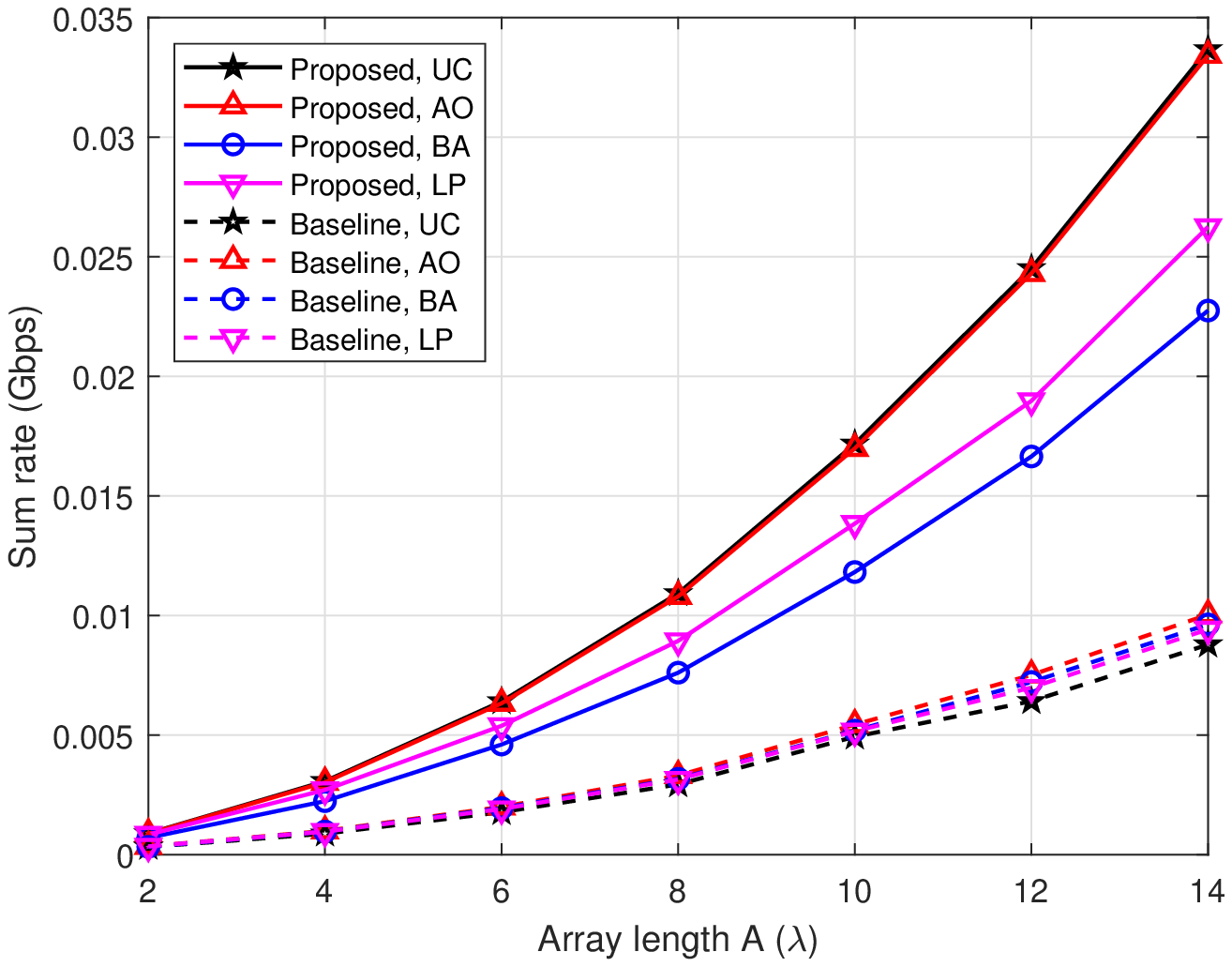}
    \caption{The comparison between the sum-rate performance of the proposed algorithms and baselines for HMA-assisted XL-MIMO communications versus the array length.}\label{fig:wideband2}
    \end{minipage}
\end{figure*}

In \figref{fig:wideband1}, we set the array length as $A = 6\lambda$ and the transmit power as $P_{\text{t}} = - 10$ dBm. We adjust the transmission bandwidth in $B \in [20,2560]$ MHz to evaluate the dual-wideband effects on the HMA-based wideband XL-MIMO uplink communications. From \figref{fig:wideband1}, we can observe that the sum rates of the proposed algorithms outperform the baselines. Additionally, the sum-rate gap between the proposed algorithms and baselines is small when the transmission bandwidth is narrow. However, the gap becomes more prominent as the transmission bandwidth increases. This phenomenon indicates that the dual-wideband effects have a worse impact on XL-MIMO communications with larger transmission bandwidths, and the proposed approach can effectively mitigate the dual-wideband effects.

In \figref{fig:wideband2}, we evaluate the dual-wideband effects on the HMA-based XL-MIMO transmission at different array sizes. We set the transmission bandwidth as $B = 600$ MHz and the transmit power as $P_{\text{t}} = -10$ dBm.
We suppose two users randomly distributed in the near field of the BS with a $16\lambda \times 16\lambda$ HMA array. In \figref{fig:wideband2}, the achievable sum rates of the proposed algorithms are higher than the baselines, indicating the effectiveness of the proposed algorithms in alleviating dual-wideband effects. We also observe that the performance gap between the proposed algorithms and baselines increases as the HMA array size becomes larger. This phenomenon indicates that the larger the HMA array size is, the more serious dual-wideband effects on the sum rate of HMA-assisted XL-MIMO communications.

%


\subsection{Near-Field Effect}

In this subsection, we compare the sum-rate performance of the proposed algorithms derived under the spherical and plane wavefront assumptions for HMA-based near-field XL-MIMO uplink communications. For the plane-wave-based channel model, we use $\tau_{u,p}$ to denote the propagation time of user $u$'s signals from the $p$th scatterer to the center of the HMA array. The propagation time of the $p$th path from user $u$ to the $(m,l)$th metamaterial element can be approximated by $ \tau_{u,p,m,l} = \tau_{u,p} +\tau_{m,l}(\theta_{u,p}, \phi_{u,p})$,
where $(\theta_{u,p}, \phi_{u,p})$ is the elevation-azimuth pair characterizing the angle of arrival \cite{2021ChannelWideband}. We denote the coordinate of the $(m, l)$th element as $(p_{m,l,x}, p_{m,l,y}, 0)$, and the time delay $\tau_{m,l}(\theta_{u,p}, \phi_{u,p})$ between the $(m, l)$th metamaterial element and the center of the HMA array as $\tau_{m,l}(\theta_{u,p}, \phi_{u,p}) = \frac{ p_{m,l,x} \sin \theta_{u,p} \cos\phi_{u,p} +  p_{m,l,y} \sin \theta_{u,p} \sin\phi_{u,p} }{c}$.
Then, the array response vector between user $u$ and the BS is modeled as
\begin{align}
  \bb_{u,p}(f) = & e^{-\jmath 2 \pi (f_c+f) \tau_{u,p}} \ \ \ntb
  & \ \ \ \cdot \left[e^{-\jmath 2 \pi (f_c+f) \tau_{1,1}(\theta_{u,p}, \phi_{u,p})}, e^{-\jmath 2 \pi (f_c+f) \tau_{1,2}(\theta_{u,p}, \phi_{u,p})}, \ldots, e^{-\jmath 2 \pi (f_c+f) \tau_{M,L}(\theta_{u,p}, \phi_{u,p})}\right].
\end{align}

\begin{figure*}[t]
    \begin{minipage}{0.49\linewidth}
    \includegraphics[width=8.2 cm]{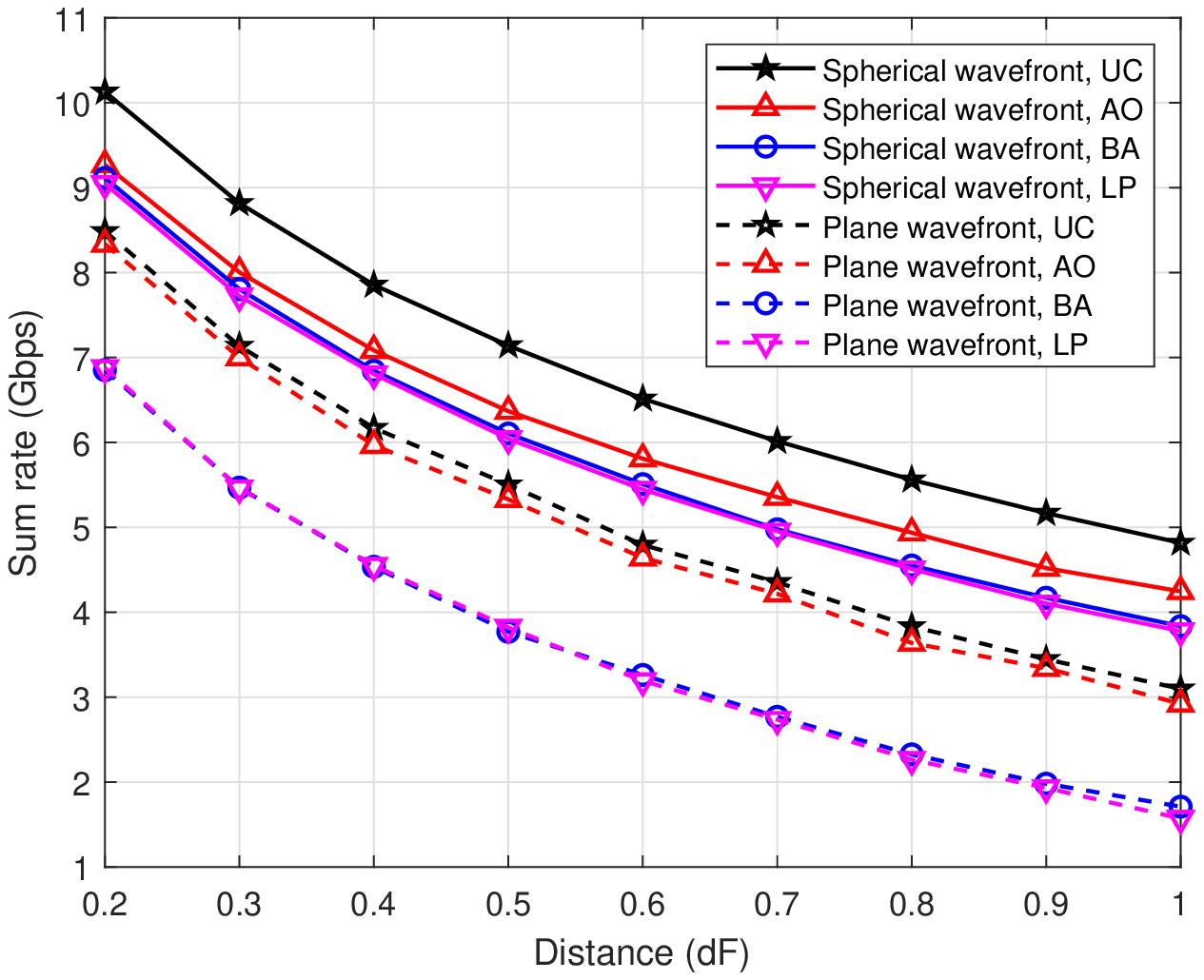}
    \caption{The comparison of the sum-rate performance between the proposed algorithms derived under spherical and plane wavefront assumptions versus the transceiver distance.}\label{fig:SW1}
    \end{minipage}
    \hfill
    \begin{minipage}{0.49\linewidth}
    \includegraphics[width=8.2 cm]{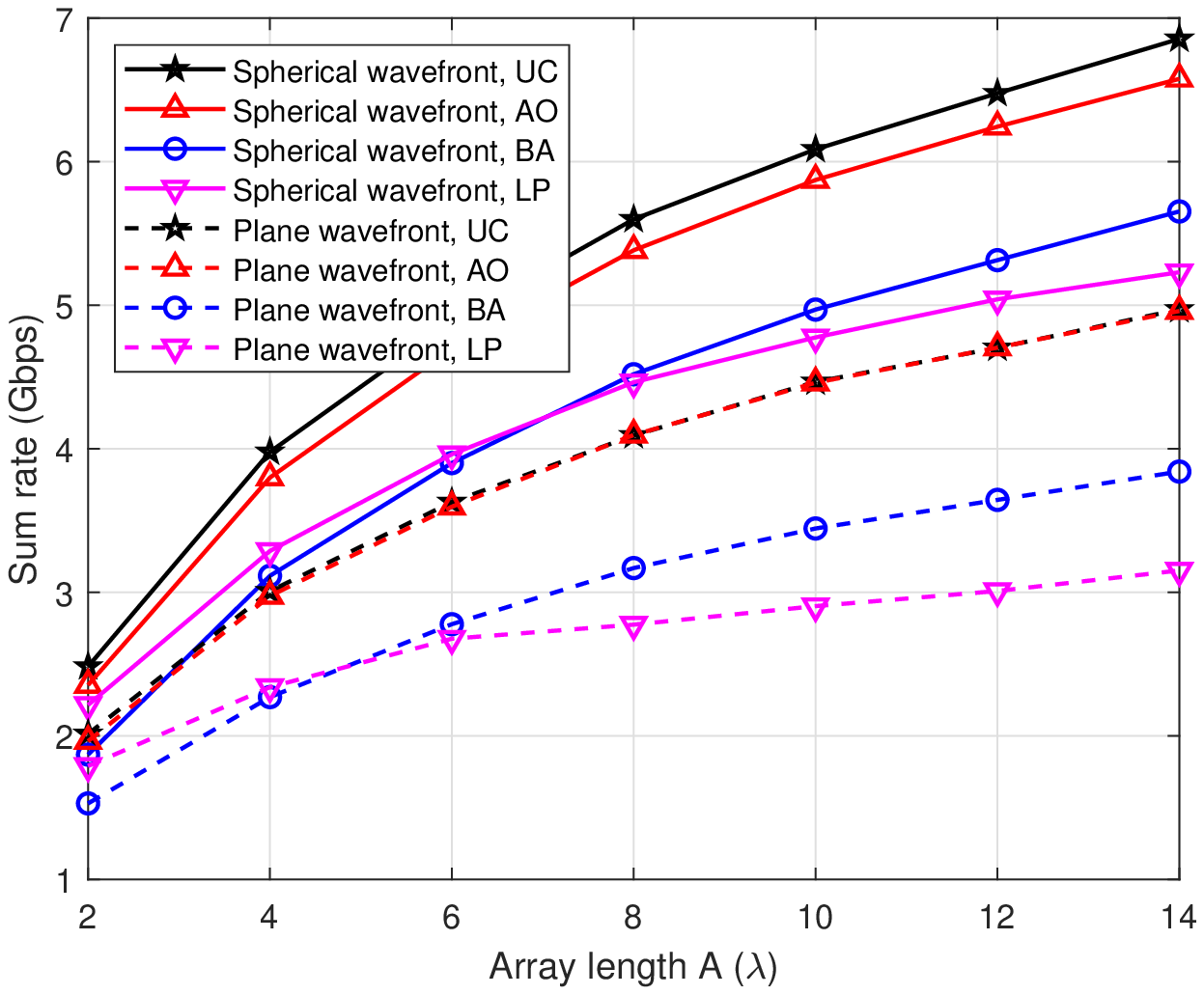}
    \caption{The comparison of the sum-rate performance between the proposed algorithms derived under spherical and plane wavefront assumptions versus the array length.}\label{fig:SW2}
    \end{minipage}
\end{figure*}


In \figref{fig:SW1}, we suppose that a single user moves along the $z$-axis in $[0.2d_{\text{F}}, d_{\text{F}}]$. We set the transmission bandwidth as $B = 600$ MHz, the array length as $A = 6\lambda$, and the transmit power as $P_{\text{t}} = 10$ dBm. From \figref{fig:SW1}, we can observe that the sum rates of the proposed algorithms derived under the spherical wavefront assumption outperform those derived under the plane wavefront assumption. Note that the angles of arrival vary significantly with the position of metamaterial elements when the transceiver distance is smaller than the Fraunhofer distance. Since the plane wave model cannot characterize different angles of arrival corresponding to different metamaterial elements, it could lead to non-negligible performance loss in near-field XL-MIMO communications. In contrast, the spherical wave model can describe the angle of arrival for each element separately, which in turn brings a significant performance gain. 

In \figref{fig:SW2}, we suppose $U = 2$ users randomly located on the $xz$-plane with angles of arrival $\theta_{u,p} \in [-\pi/4, \pi/4]$ in the near-field region of the BS with a $16\lambda \times 16\lambda$ HMA array. We set the transmission bandwidth as $B = 600$ MHz. It can be observed again that the sum rates of the proposed algorithms under the spherical wavefront assumption are better than those under the plane wavefront assumption. In addition, the sum-rate gap between the proposed spherical wavefront and the conventional plane wavefront cases becomes more prominent when the array length increases. Hence, our proposed algorithms under the spherical wavefront assumption can effectively alleviate the negative near-field effect in the HMA-assisted XL-MIMO uplink communications in the near-field region.

%

\subsection{Impact of Number of Users}
\begin{figure}[t]
\centering
\includegraphics[width=0.7\textwidth]{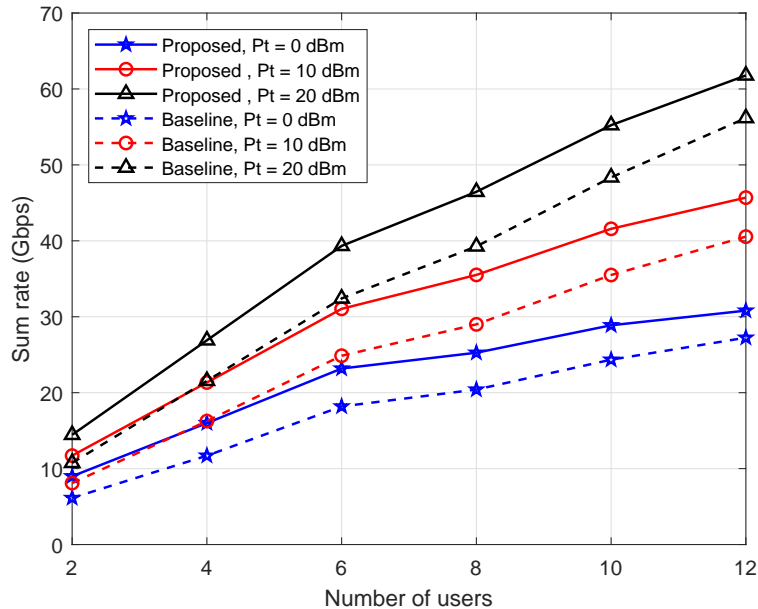}
\caption{The sum-rate performance of the proposed algorithms and baselines versus the number of users.}
\label{fig:UserNum}
\end{figure}
In \figref{fig:UserNum}, we study a more general case with different numbers of users randomly distributed in the $xz$-plane with the angle of arrival $\theta_{u,p} \in [-\pi/4, \pi/4]$ in the near-field region. We use ``baseline" to represent the sum-rate performance of the conventional beam combining design that utilizes the plane wave model and neglects dual-wideband effects in the HMA-assisted XL-MIMO uplink communications. We set $A = 10 \lambda$, $B = 600$ MHz, and $P_{\text{t}} = 0$, $10$, and $20$ dBm. Since the curves of four sets of HMA weights scale similarly, we present the sum-rate performance of the Lorentzian-constrained phase case and omit the other cases without loss of generality. From \figref{fig:UserNum}, we can observe that the achievable sum rates of the proposed algorithms outperform the baselines. This phenomenon illustrates in a more general way that our proposed algorithms can effectively alleviate the
sum-rate loss caused by the near-field and dual-wideband effects in HMA-assisted XL-MIMO uplink systems.

\subsection{Comparison with Other Schemes}
\begin{figure}[t]
\centering
\includegraphics[width=0.7\textwidth ]{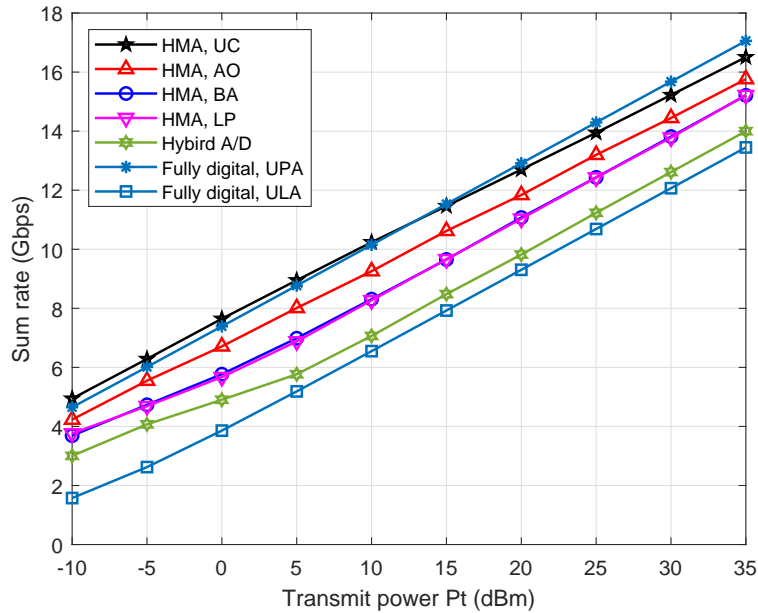}
\caption{The sum-rate performance of the proposed HMA-based beam combining and two conventional beam combining architectures versus the transmit power $P_\text{t}$.}
\label{fig:Hybrid}
\end{figure}

In \figref{fig:Hybrid}, we compare the sum-rate performance of three architectures: the proposed HMA-assisted, conventional fully-connected hybrid A/D, and conventional fully digital architectures. Besides, we consider two specific fully digital architectures. In one case, the fully digital architecture has the same array aperture as the HMA-assisted one, and the antenna elements are arranged in a uniform planar array (UPA). In the other case, the number of conventional antennas is the same as that of microstrips in the HMA-assisted one, and the antennas are arranged in a uniform linear array (ULA). Note that our proposed algorithmic framework can also be applied to design the hybrid A/D beam combining. Specifically, since the entries of the RF combining matrix are all unit modulus in the hybrid A/D architecture, the corresponding maximization problem is only constrained by $|(\bW_{\text{RF}})_{i,j}| = 1$, $\forall i, j$, where $\bW_{\text{RF}}$ represents the RF combining matrix \cite{2020Hybrid}. In addition, we adopt the approach provided in \cite{2014Spatially} to design the fully digital beam combiners.

We set $A = 6 \lambda$, $B = 600$ MHz, and $U=2$. In \figref{fig:Hybrid}, the HMA-based beam combining achieves a higher sum rate than the conventional hybrid A/D architecture. This is because HMA-based arrays can accommodate more antenna elements than the conventional antenna arrays, compensating for the lower gain of a single metamaterial element than a single conventional patch antenna. Besides, all four kinds of HMA-based beam combining, especially the unconstrained weight case, achieve lower sum-rate performance than the conventional fully digital one with an UPA. This performance disadvantage is mainly due to the significantly reduced demand on the number of RF chains in the HMA-assisted BS, which reduces the signal processing capability in the digital domain. However, it can be observed that HMA-based beam combining can achieve much higher sum rates than the conventional fully digital one with the same number of RF chains. Note that a metamaterial element is cost-competitive, power-efficient, and sub-wavelength \cite{2021Dynamic6G}. Therefore, increasing metamaterial elements on each microstrip can bring a significant sum-rate gain with acceptably additional power consumption. Meanwhile, HMA-based beam combining is much more energy-efficient than fully digital and hybrid A/D ones \cite{2021DMAsEE}. Thus, it is expected to realize HMA-assisted XL-MIMO systems with much better EE and sum rate than traditional full-digital systems by appropriately increasing the number of metamaterial elements on each waveguide.

%
\subsection{Impact of Mutual Coupling}

Holographic MIMO integrates as many antennas as possible into a limited area to increase the system capacity \cite{2020Holographic}. However, when the antenna elements are closely spaced, the mutual
coupling between the elements becomes significant and
may reduce the system performance \cite{2012SunCapacity}. 
In this subsection, we evaluate the achievable sum rates of the proposed HMA-assisted and conventional XL-MIMO systems in the presence of mutual coupling.

For the case incorporating the mutual coupling, the post-process receiving signal in \eqref{eq:received_signal} can be slightly modified as $ \bz_s = \bW_s^H \bQ \bH_s \bC \bG_s\bx_s + \bW_s^H \bQ \bH_s \bn_s$, where $\bC \in \C^{N_{\text{R}} \times N_{\text{R}} }$ is the mutual coupling matrix given by \cite{2005AntenTheory}
\begin{align}\label{eq:C}
  \bC = \left( Z_{\text{A}} + Z_{\text{L}} \right) \left( \bZ +  Z_{\text{L}} \bI \right)^{-1}.
\end{align}
In \eqref{eq:C}, $Z_{\text{A}}$ is the antenna impedance and $Z_{\text{L}}$ is the load impedance. Both $Z_{\text{A}}$ and $Z_{\text{L}}$ are fixed to 50 Ohms \cite{2003clerckxmutual}. The mutual impedance matrix $\bZ$ is formulated as
    \begin{align}\label{eq:mutual_impedance}
        \begin{pmatrix}
        Z_{\text{A}} & Z_{12} & \ldots & Z_{1 N_{\text{R}}}\\
          Z_{21} & Z_{\text{A}} & \ldots &  Z_{2  N_{\text{R}}}  \\
           \vdots &  \vdots &  \ldots & \vdots  \\
          Z_{N_{\text{R}} 1} & Z_{N_{\text{R}} 2} &  \ldots & Z_{\text{A}}
         \end{pmatrix}.
    \end{align}
We consider a set of side-by-side wire dipoles of length $L_{\text{d}}$ regularly arranged on an UPA. Typically, the conventional antennas are half-wavelength dipoles \cite{2003clerckxmutual} while the metamaterial elements are dipoles of $1/32$ wavelength \cite{2021GradoniEnd}. The off-diagonal entry of the mutual impedance matrix $\bZ$ is given by \cite{2005AntenTheory}
  \begin{align}\label{eq:Zmn}
        Z_{mn} = & 30\left[ 2 \cC_i\left(\kappa d_{mn} \right) - \cC_i\left(\kappa \left( \sqrt{d_{mn}^2 + L_{\text{d}}^2} + L_{\text{d}} \right) \right) - \cC_i\left(\kappa \left( \sqrt{d_{mn}^2 + L_{\text{d}}^2} - L_{\text{d}} \right) \right) \right] \ntb
        -&30\jmath \left[ 2 \cS_i\left(\kappa d_{mn}\right) - \cS_i\left(\kappa \left( \sqrt{d_{mn}^2 + L_{\text{d}}^2} + L_{\text{d}} \right) \right) - \cS_i\left(\kappa \left( \sqrt{d_{mn}^2 + L_{\text{d}}^2} - L_{\text{d}} \right) \right) \right],
    \end{align}
where $\kappa = 2\pi/\lambda$ is the wavenumber and $d_{mn}$ is the distance between two dipoles. Besides, $\cC_i(\cdot)$ and $\cS_i(\cdot)$ are cosine and sine integrals respectively given by
\begin{subequations}
  \begin{align}
     \cC_i(u) = \int_{\infty}^{u} \frac{\cos x}{x}dx, \\
     \cS_i(u) = \int_{0}^{u} \frac{\sin x}{x}dx.
  \end{align}
\end{subequations}
The proposed algorithmic framework can be applied to address  the sum-rate maximization problem of the considered XL-MIMO systems with the existence of mutual coupling by treating $\bC\bG_s$ as the equivalent channel matrix.

\begin{figure}
\centering
\subfloat[]{\centering\includegraphics[width=0.48\textwidth]{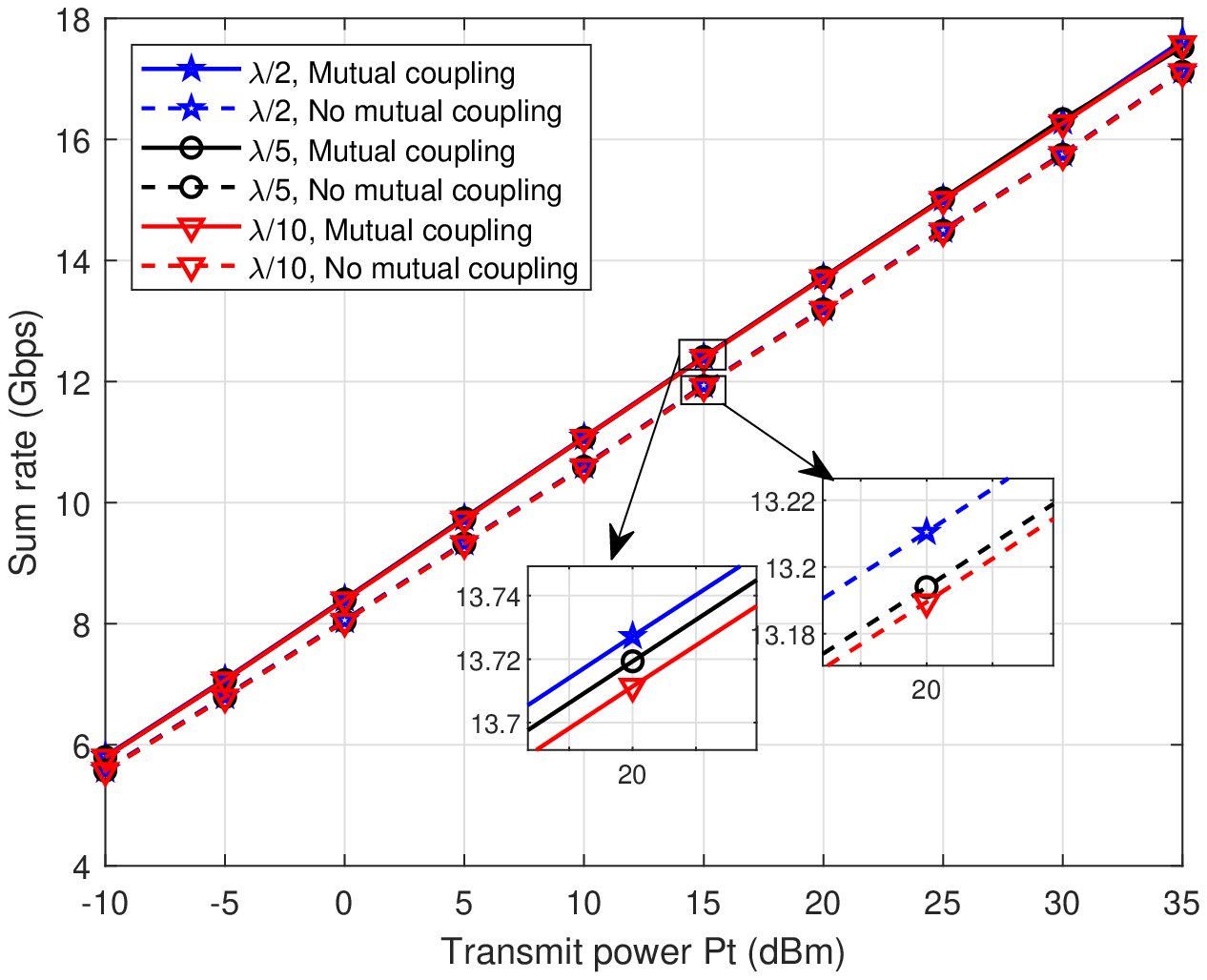}}\hfill
\subfloat[]{\centering\includegraphics[width=0.48\textwidth]{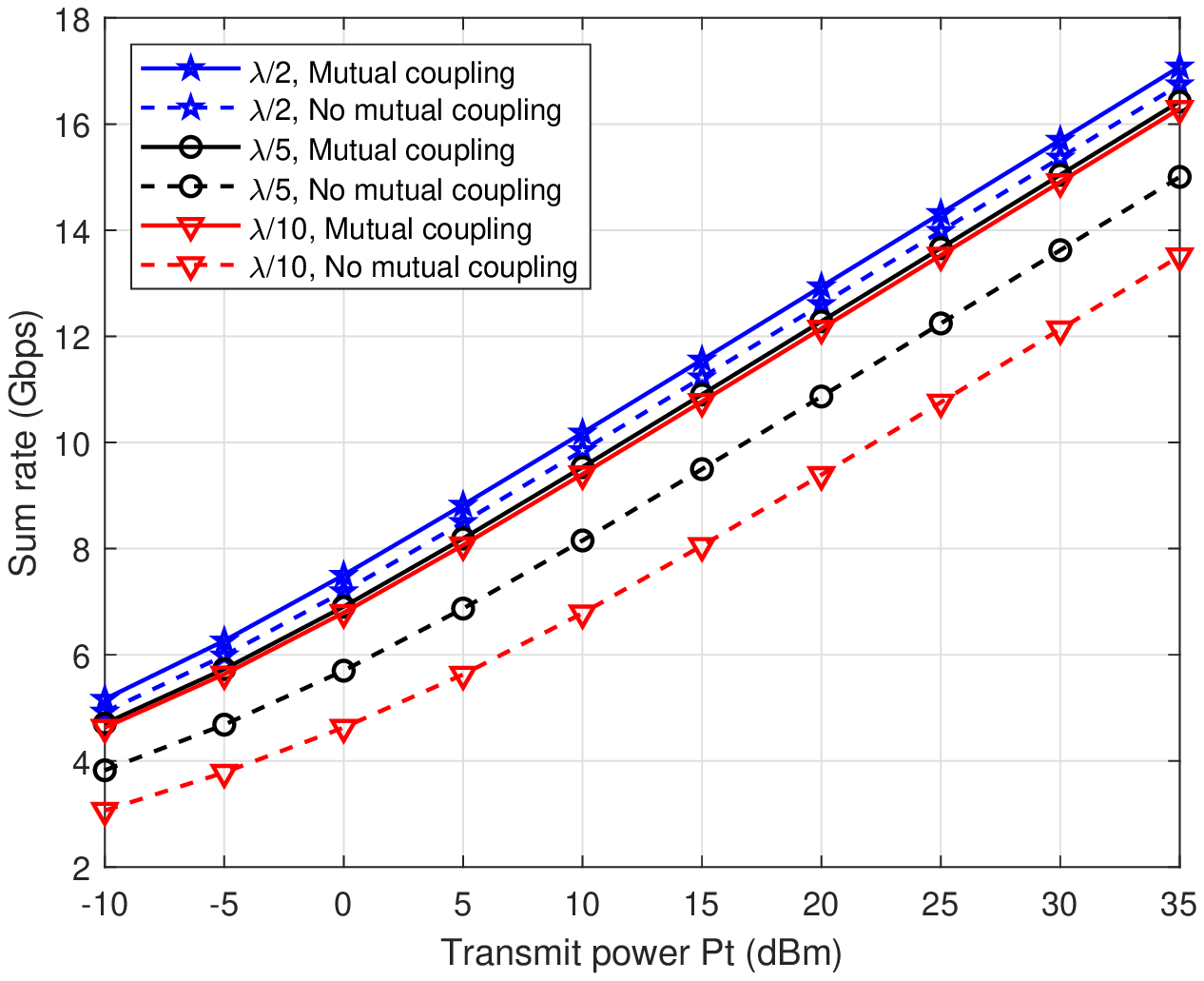}}
\caption{Impact of mutual coupling on the sum-rate performance: (a) HMA-assisted XL-MIMO transmission; (b) conventional-antenna-assisted XL-MIMO transmission.}
\label{fig:couplingCom}
\end{figure}

In \figref{fig:couplingCom}, ``Mutual coupling" denotes the sum-rate maximization design incorporating the mutual coupling between antenna elements, while ``No mutual coupling" indicates the opposite case where the design neglects the mutual coupling. We set $A = 6 \lambda$, $B = 600$ MHz, and $U=2$. \figref{fig:couplingCom}(a) characterizes the impact of mutual coupling in the proposed HMA-assisted XL-MIMO communications. It can be observed that the mutual coupling brings a slight performance loss in the sum rate of the proposed HMA-assisted system, which is negligible. Meanwhile, the negative effect of mutual coupling does not increase with the further decrease of antenna intervals. This phenomenon benefits from the very short dipole length compared to the antenna interval in the HMA array. On the contrary, \figref{fig:couplingCom}(b) shows that the mutual coupling leads to great loss in the sum-rate performance when the antenna interval is smaller than $\lambda /2$ in conventional-antenna-assisted XL-MIMO communications. Besides, the negative influence is further increased by narrowing the antenna interval in the conventional-antenna array. Therefore, \figref{fig:couplingCom} indicates that mutual coupling has little effect on the sum-rate performance of HMA-assisted systems, while mutual coupling should be taken into account in conventional-antenna-assisted systems. This feature makes metasurface antennas more conducive to the development of holographic MIMO compared to the costly and seriously-coupled conventional antennas.

\section{Conclusion}\label{sec:Conclusion}
In this paper, we studied the HMA-based wideband XL-MIMO uplink beam combining design in the near-field region with the goal of sum-rate maximization. Firstly, we introduced a spherical-wave-based channel model that
simultaneously takes into account both the near-field and dual-wideband effects. Based on this model, we proposed an algorithmic framework to alleviate the near-field and dual-wideband effects for HMA-based XL-MIMO beam combining, including the MMSE method, AO method, matrix vectorization rule, and MM method. Numerical results showcased that the proposed algorithms can effectively reduce the near-field and dual-wideband effects on the sum-rate performance of HMA-assisted XL-MIMO systems. Moreover, for the same array aperture size, the HMA-based beam combining could achieve a better sum rate than the hybrid A/D one, as well as a comparable sum rate to the conventional fully digital one. In addition, it was demonstrated that the sum-rate performance of the HMA-assisted system is almost unaffected by mutual coupling, even when the antenna interval is sub-wavelength.

\begin{appendices}
\section{Proof of \eqref{eq:tr2}, \eqref{eq:tr1}, and \eqref{eq:tr3}}\label{sec:Appendix_A}
 Eq. \eqref{eq:tr2} can be proved by $\tr{\bQ\bC_s} = \sum_{k=1}^{M}\sum_{l=1}^L (\bQ)_{m,(m-1)L+l} (\bC_s)_{(m-1)L+l,m} = \bq^T \bc_s$ and $\tr{\bQ^H \bC_s^H} = \tr{\bC_s^H \bQ^H} = \sum_{k=1}^{M}\sum_{l=1}^L (\bC_s)_{(m-1)L+l,m}^{*} (\bQ)_{m,(m-1)L+l}^{*} = \bc_s^H \bq^{*}$. To prove \eqref{eq:tr1}, we use the properties $\tr{\bR^T \bT} = \text{vec}(\bR)^T \text{vec}(\bT)$, $\tr{\bR \bT} = \tr{\bR \bT}$ and $\text{vec}(\bR \bX \bT) = (\bT^T \otimes \bR) \text{vec}(\bX)$. Then, we have
    \begin{align}
        \tr{ \bQ^H \bB_s \bQ \bA_s }  &= \tr{ \bQ \bA_s \bQ^H \bB_s} = \text{vec}(\bQ^T)^T \text{vec}(\bA_s \bQ^H \bB_s)\ntb
        & = \text{vec}(\bQ^T)^T (\bB_s^T \otimes \bA_s) \text{vec}(\bQ^H)  = \text{vec}(\bQ^T)^H (\bB_s \otimes \bA_s^T) \text{vec}(\bQ^T).
    \end{align}
For brevity, we define $\bar{\bq} = \text{vec}(\bQ^T) = [ q_{1,1}, q_{1,2}, \ldots, q_{1,L},\underbrace{ 0, 0, \ldots, 0}_{N_R}, q_{2,L+1}, \ldots, q_{M,N_R} ]^T $ and $\bq = [q_{1,1}, q_{1,2}, \ldots, q_{m,(m-1)L+l}, \ldots, q_{M,ML}]^T$, where $\bq$ removes the zero entries in $\bar{\bq}$. Then, we expand $\text{vec}(\bQ^T)^H (\bB_s \otimes \bA_s^T) \text{vec}(\bQ^T)$ as
    \begin{align}
    \text{vec}(\bQ^T)^H (\bB_s \otimes \bA_s^T) \text{vec}(\bQ^T) &=
     \bar{\bq}^H
     \begin{pmatrix}
           b_{1,1}\bA_s^T & b_{1,2}\bA_s^T & \ldots & b_{1,M}\bA_s^T \\
          b_{2,1}\bA_s^T & b_{2,2}\bA_s^T & \ldots &  b_{2,M}\bA_s^T     \\
           \vdots &  \vdots &  \ldots & \vdots  \\
          b_{M,1}\bA_s^T & b_{M,2}\bA_s^T &  \ldots & b_{M,M}\bA_s^T
         \end{pmatrix} \bar{\bq} \ntb
          &= \bq^H \bD \bq.\label{eq:matrix_vec}
    \end{align}
In \eqref{eq:matrix_vec}, $\bD$ can be represented as
\begin{align}
   &\bD = \ntb
   &\begin{small}
    \begin{pmatrix}
    b_{1,1}\bA_{s,1\sim L,1\sim L}^T & b_{1,2}\bA_{s,L+1\sim 2L, 1\sim L}^T & \ldots &            b_{1,M}\bA_{s,(M-1)L+1\sim N_R, 1\sim L}^T \\
    b_{2,1}\bA_{s,1\sim L, L+1\sim 2L}^T & b_{2,2}\bA_{s,L+1\sim 2L, L+1\sim 2L}^T & \ldots &              b_{2,M}\bA_{s,(M-1)L+1\sim N_R, L+1\sim 2L}^T     \\
    \vdots &  \vdots &  \ldots & \vdots  \\
    b_{M,1}\bA_{s,1\sim L, (M-1)L+1\sim N_R}^T & b_{M,2}\bA_{s,L+1\sim 2L, (M-1)L+1\sim N_R}^T &  \ldots &              b_{M,M}\bA_{s, (M-1)L+1\sim N_R, (M-1)L+1\sim N_R}^T
    \end{pmatrix} \end{small}
   \ntb
   & = \left( \bB_s \otimes \bN_L\right) \odot \bA_s^T,
\end{align}
where $\bN_L$ is a all-one matrix. Hence, we obtain $\tr{ \bQ^H \bB_s \bQ \bA_s } = \bq^H \left( \bB_s \otimes \bN_L\right) \odot \bA_s^T \bq$ and prove Eq. \eqref{eq:tr1}. The proof of \eqref{eq:tr3} can be obtained in the similar way.

\end{appendices}

\end{document}